\documentclass[sigconf]{acmart}

\usepackage{comment}
\usepackage{tabularray}
\usepackage{tabularx}
\usepackage{graphicx}
\usepackage{rotating}
\usepackage{enumitem}
\usepackage{booktabs}
\usepackage[table]{xcolor}
\usepackage{tikz}
\usepackage{subcaption}
\usepackage{xcolor}
\usepackage{soul}

\definecolor{lightpurple}{RGB}{220,200,255}

\newcommand{\toolname}{EvoGraph}

\usepackage[most]{tcolorbox}

\newtcolorbox{guidelinebox}{
  colback=gray!15,
  colframe=gray!40,
  boxrule=0.3pt,
  arc=2pt,
  left=4pt,
  right=4pt,
  top=3pt,
  bottom=3pt,
  before skip=4pt,
  after skip=0pt
}

\newcommand{\designguideline}[1]{
\begin{guidelinebox}
\textbf{#1}
\end{guidelinebox}
}

\usepackage{ifthen}
\usepackage{xstring}

\usepackage{pgf}

\newcommand{\pvalue}[1]{
  \ensuremath{
    \pgfmathparse{#1 < 0.001}
    \ifnum\pgfmathresult=1
      p < 0.001
    \else
      \pgfmathparse{#1 < 0.01}
      \ifnum\pgfmathresult=1
        p < 0.01
      \else
        \pgfmathparse{#1 < 0.05}
        \ifnum\pgfmathresult=1
          p < 0.05
        \else
          p = \pgfmathprintnumber[fixed,precision=3]{#1}
        \fi
      \fi
    \fi
  }
}

\newcommand*\circled[1]{\raisebox{.5pt}{\textcircled{\raisebox{-.7pt} {#1}}}}

\newcommand{\participantquote}[1]{
  {\small\textsf{``#1''}}
}
\newcommand{\participantquoteattributed}[2]{
  {\small\textsf{``#1''}(\textit{#2})}
}
\newcommand{\sig}[1]{\ifmmode^{#1}\else$^{#1}$\fi}

\copyrightyear{2026}
\acmYear{2026}
\setcopyright{cc}
\setcctype{by-sa}
\acmConference{Conference'26}{XX, XX}

\begin{document}
\title{Choose Your Own Adventure: Non-Linear AI-Assisted Programming with EvoGraph}

\author{Vassilios Exarhakos}
\affiliation{
  \institution{McGill University}
  \city{Montreal}
  \state{Quebec}
  \country{Canada}
}
\email{vassilios.exarhakos@mail.mcgill.ca}

\author{Jinghui Cheng}
\affiliation{
  \institution{Polytechnique Montreal}
  \city{Montreal}
  \state{Quebec}
  \country{Canada}
}
\email{jinghui.cheng@polymtl.ca}

\author{Jin L.C. Guo}
\affiliation{
  \institution{McGill University}
  \city{Montreal}
  \state{Quebec}
  \country{Canada}
}
\email{jguo@cs.mcgill.ca}

\begin{abstract}
Current AI-assisted programming tools are predominantly linear and chat-based, which deviates from the iterative and branching nature of programming itself. Our preliminary study with developers using AI assistants suggested that they often struggle to explore alternatives, manage prompting sequences, and trace changes. Informed by these insights, we created \toolname{}, an IDE plugin that integrates AI interactions and code changes as a lightweight and interactive development graph. \toolname{} automatically records a branching AI-assisted coding history and allows developers to manipulate the graph to compare, merge, and revisit prior collaborative AI programming states. Our user study with 20 participants revealed that \toolname{} addressed developers' challenges identified in our preliminary study while imposing lower cognitive load. Participants also found the graph-based representation supported safe exploration, efficient iteration, and reflection on AI-generated changes. Our work highlights design opportunities for tools to help developers make sense of and act on their problem-solving progress in the emerging AI-mediated programming context.

\end{abstract}


\maketitle

\section{Introduction}

As large language models (LLMs) have grown more capable~\cite{swebench}, AI programming assistants, such as GitHub Copilot~\cite{Copilot2025}, Cursor~\cite{Cursor2025}, and Claude Code~\cite{ClaudeCode2025}, are transforming how developers search for information, write code, and resolve bugs~\cite{liang2023largescalesurveyusabilityai, Barke_Grounded_Copilot}.
Developers are increasingly adopting these tools, with the goal of focusing more on higher-level, creative, and strategic aspects of software design and accelerating the implementation process. Despite the promises of AI-programming assistants liberating their work and increasing their productivity, however, many developers cast doubt on AI adoption or only obtain limited gains from it~\cite{StackOverflowSurvey2025}.

An important reason for this is that the linear, chat-based modality common for LLM-based interactions does not align well with the inherently non-linear nature of programming. Yet, most state-of-the-art tools operate by embedding such a chat interface inside the integrated development environment (IDE). This integration provides several benefits, such as allowing AI to access the project repository as context and minimizing tool switching for developers~\cite{TheProgrammersAssistant, NamMHVM24_LLM_for_Code_Understanding}. However,  
this linear chat modality does not account for activities such as iterative exploration, working in parallel paths, and backtracking, all of which are essential to effective problem-solving in software development. While version control systems familiar to developers can mitigate the issue to some extent, they are cumbersome to use and do not meet the fast-evolving pace of AI-enabled exploration.

To address this, we propose \textbf{\toolname{}} to support AI-assisted programming that is non-linear in nature. To identify the concrete challenges faced by developers in their current AI-mediated workflows, we started with an interview study with eight developers who had a range of experience levels. We found that developers lack practical solutions to explore alternative designs, compare them, and converge them to the ideal outcomes. Moreover, the AI assistant cannot fully understand the developers' intents; this results in long prompting sequences that developers cannot keep track of, which disrupt the context window for the AI assistant. Finally, reviewing the codebase is difficult when it becomes a melange of changes layered by developers and the AI assistant, attributed to different prompts. 
Built on those findings, we proposed three guidelines for designing AI-assisted programming tools. 

Following those guidelines, we designed and implemented \toolname{}, an IDE plugin that surfaces the history of code and AI-prompting simultaneously as a lightweight and interactive development graph to help developers easily make sense of and act on their prior collaborative work with AI. \toolname{} captures the prompting sequences and code changes automatically and presents them as nodes on the graph. Developers can directly revisit individual nodes to locate earlier prompting context or codebase status and branch out in parallel explorations without losing their original work; they can also compare multiple nodes and merge the corresponding alternative paths. Any changes in the codebase are tracked back to their originating prompting context for developer reviews.

To evaluate the impact of \toolname{} on the AI-assisted programming process, we conducted a task-based, within-subjects user study with 20 participants, where they performed two programming tasks using \toolname{} and a baseline AI-assisted programming interface in a counterbalanced order.
Our participants perceived that \toolname{} better supports them in exploring alternative solutions, managing prompting interactions, and tracking AI-generated changes, while imposing lower cognitive load. Participants also reported that the graph-based representation enabled safe exploration, supported efficient iteration, and encouraged reflection on AI-generated changes. 
Overall, our work highlights opportunities for richer versioning mechanisms, more navigable representations of prompting histories, and more expressive forms of provenance tracking for future AI-assisted programming tools. Considering the ever-evolving landscape of AI-assisted programming, we also provide implications on how to leverage and extend features of \toolname{} to emerging AI-assisted development contexts, to oversee autonomous AI systems, enhance accountability and traceability, and support AI-mediated collaborative programming.
\section{Related Work}

\subsection{Programming as Problem Solving}

How programmers understand and interact with the code plays a significant role in their problem-solving process. \citet{10.1145/1287624.1287673} found that the time and effort spent on the code by the developers and the stability of the code itself are direct indicators of the developer's knowledge of the program. This might explain why, in exploratory settings, developers often experience difficulty managing the frequently changing codebase, especially in notebook environments, where the flexible code structure and execution order afford quick code sketching but reduce its stability~\cite{kery_variolite,kery_story_in_the_notebook}.

One way to support programming activity is to enhance information seeking and navigation in the codebase. Through a combined EEG and eye-tracking study, \citet{10.1145/3540250.3549084} found that the efficiency of program comprehension is correlated with targeted focusing, code skipping, and lower cognitive load. With augmented information about code behavior (such as the relationship between code elements, or code and its output), programmers make fewer mistakes~\cite{10.1145/3025453.3025645} and better pinpoint specific code that contributes to their understanding of the program behavior~\cite{interlink,10.1145/2047196.2047225}. 
Another way to support programming is to consider the evolving nature of the codebase, supporting visualizing and interacting with the code change history. Prior work has looked at micro-versioning, which emphasizes lightweight, frequent capture of intermediate states, as a way of tracking changes in a rapidly evolving codebase~\cite{mikami_microversioning, kery_variolite, Spellburst, managingMesses}. 
For example, in data science, \citet{kery_variolite} developed Variolite, a lightweight versioning tool that enables data scientists to capture and switch between alternative code variants within the notebook environment.

Our work builds on prior approaches to program comprehension and evolution by extending them to AI-assisted workflows. With the introduction of AI assistants as intermediaries in the programming process, developers are no longer interacting solely with code but also with evolving AI-generated suggestions and interaction histories. This mediation fundamentally changes how information is sought, interpreted, and validated, introducing new challenges in understanding how code is produced and how it evolves over time.

\subsection{AI-Assisted Programming Workflow}

AI-assisted programming involves fluid, iterative workflows where developers alternate between acceleration and exploration~\cite{Barke_Grounded_Copilot}, using AI both to speed up implementation~\cite{WideningGap,liang2023largescalesurveyusabilityai} and to brainstorm and experiment with new ideas at a higher level of abstraction~\cite{IBM_Examining_Use_Impact_AI_on_Developer_Productivity_Experience,bicycle, SERGEYUK_AI_assistants_in_practice, BeyondCodeGenerationGPT}.
However, in this workflow, developers have to manage long, iterative interactions as well as understand both the generated code and how it evolves.

Developers often struggle to formulate effective prompts for AI coding assistants~\cite{liang2023largescalesurveyusabilityai}.
To manage complexity, developers decompose tasks into smaller subtasks and iteratively construct solutions through sequences of prompts~\cite{liang2023largescalesurveyusabilityai,2022_Usability_of_code_generation_tools_Expectation_vs_Experience}. 
However, this workflow may steer developers toward incorrect or suboptimal solution paths~\cite{WideningGap, Its_weird_it_knows_what_I_want}, as AI outputs may appear plausible, but fail to meet functional or non-functional requirements~\cite{liang2023largescalesurveyusabilityai}. 
These challenges are further compounded by limitations in current tools, where previous explorations are simply treated as transient history that cannot be effectively navigated and recovered from~\cite{2022_Usability_of_code_generation_tools_Expectation_vs_Experience}.

As a result, developers must continuously adapt their reliance on the assistant, as trust remains situational and shaped by the assistant’s perceived ability and reliability~\cite{designingTrust}.
The speed afforded by these tools can come at the cost of code quality and comprehension. Developers may engage in \emph{comprehension outsourcing}, accepting AI-generated code without fully understanding it~\cite{NamMHVM24_LLM_for_Code_Understanding}, and recent work suggests that while AI assistance can increase development speed, it may also introduce more warnings and increase code complexity~\cite{he2025doesaiassistedcodingdeliver}.
Prior work has emphasized the importance of making both human and AI contributions visible to support collaboration, code review, and attribution~\cite{ExplainabilityScenario,IBM_Examining_Use_Impact_AI_on_Developer_Productivity_Experience}. 
Providing such transparency can improve developers’ ability to validate and repair code~\cite{eyeTracking}. 

Overall, prior work highlights challenges in steering and validating iterative AI interactions, as well as risks of overreliance. Our tool builds on these insights by providing mechanisms for inspecting and managing interaction history at a fine-grained level and supporting reasoning about AI and human contributions.

\subsection{Tools Supporting AI-Assisted Programming}

A growing body of work has explored how to better support developers in working with LLM-based programming assistants. Several systems aim to bridge the gap between natural language prompts and executable code by introducing more structured interfaces. 
This includes approaches that allow developers to guide and refine outputs~\cite{BISCUIT, designingTrust}, proactively anticipate developer needs~\cite{Pu_Assistance_or_Disruption,NeedHelp}, and structure interactions across levels of task decomposition~\cite{coladder}.
For supporting alternative exploration, \citet{Zamfirescu_Pereira_2025_LLM-supported_Exploration_Program_Design_Space} developed an IDE that generates alternative design options, tracks design decisions, and highlights implicit choices. This work shifts the focus to the requirements and design activities and allows developers to provide instructions at a higher level of abstraction.

Prior work has also explored tools to support understanding of AI-generated code, such as Ivie, providing real-time contextual explanations to improve comprehension without disrupting workflow~\cite{ivie}, and Trailblazer, offering structured narratives with summaries, code snippets, and replayable walkthroughs to explain both outcomes and reasoning~\cite{programTraces}.
Similarly, Meta-Manager tracks code provenance across sources, enabling developers to understand unfamiliar codebases~\cite{metaManager}.

Finally, several tools aim to support the verification of AI-generated code. 
Live programming interfaces provide real-time feedback to help developers preview and validate outputs~\cite{liveProgramming}, while interactive workflows such as TICODER use test-driven loops to iteratively refine code based on execution and user feedback~\cite{TICODER}.
Other work explores uncertainty-aware interfaces, such as highlighting tokens that are likely to require edits, which can guide developer attention and support more targeted validation~\cite{UncertaintyHighlighting, PerfectionNotRequired}. 

Our work aims to support comprehension and validation broadly by providing an integrated understanding of both code and chat history, allowing developers to navigate and compare iterations in the context of AI-assisted programming.
\section{Preliminary Interview Study}
\label{sec:preliminary_study}

We conducted a preliminary interview study with developers experienced in AI-assisted programming to refine our understanding of their needs and to distill concrete system design guidelines.

\subsection{Study Methods}
The study was approved by the research ethics board of all involved institutions. Through local computer science-related Slack channels and personal networks, we recruited eight participants (aged 20–29 years, five males, three females) with a range of experience levels in software development: one had less than 1 year, one had 1–3 years, four had 3–5 years, and two had 5–10 years of experience. Four reported using AI-powered programming tools daily; two, a few times per week; and the remaining two, a few times per month. 

Each interview session, conducted remotely over Microsoft Teams, lasted approximately 30 minutes, in which participants discussed their current experience, processes, and challenges using AI-powered programming tools. We then conducted a thematic analysis~\cite{braun_thematic_2021} on the transcribed recording to identify recurring themes across participants' experiences and derive design implications.

\subsection{Results}

\paragraph{Overall process} Despite using different tools, our participants described a similar process for AI-assisted programming: \textbf{an iteration of multiple prompting, verification, and problem-fixing cycles}. For each stage, they mentioned a wide variance of activities ranging from  \textit{carefully planned executions} to \textit{spontaneous and reactive actions}. For example, when confirming functionality and correctness of the generated code during the \textit{verification} stage, P1 relied heavily on existing unit tests to systematically verify that outputs of AI-generated code matched expectations. P7, on the other hand, validated the AI's results (front-end features in this case) by simply \participantquote{clicking through the website, seeing if the way it reacts is how I want it to.} Some applied strategies that combined automated checks (e.g., using linters [P8]) and manual inspection. 

Regardless of the concrete strategies, the iterative nature of AI-assisted programming has caused significant challenges for the participants. They have to frequently go back-and-forth with the AI assistants, prompting to clarify their intention and correct the AI assistants' behavior, and sometimes abandoning it entirely, considering it \participantquoteattributed{useless to continue the chat}{P3}. Below, we describe three emerging themes from participants' discussion on the challenges and needs when performing AI-assisted programming.

\subsubsection{Managing Exploration History}
\label{subsubsec:challenge_manage_history}

To manage the exploratory nature of AI-assisted programming, some participants used established version control systems, such as Git, to manage AI-generated changes. However, they found that \textbf{the granularity of the traditional version control commits does not align with the small, frequent, and sometimes temporary changes made when using an AI assistant}.
For example, although P4 tried to make a commit at each change the LLM makes, they found it \participantquote{annoying} and often forgot to do so.
Due to their inability to manage AI changes on a fine-grained scale, when taking a misstep, participants often had to remove all code generated since their previous commit. 
For instance, P7 described situations in later stages of development where previously working AI-generated code became destabilized by subsequent prompts: when there were \participantquote{a lot of moving parts} and the LLM was \participantquote{not keeping context of the other elements on the page}, new changes would unintentionally disrupt existing functionality. In these cases, they were forced to roll back the accumulated AI changes entirely, losing previously working code.

Alternatively, some participants \textbf{turned to manual strategies or external tools to manage code history but found them cumbersome and not scalable}. For example, P1 saved pre-existing code snippets into a local Notepad file before prompting the model for alternative implementations. 
On the other hand, P7 described scrolling through the AI chat to locate and undo recent edits; P3 and P4 similarly mentioned asking the AI assistant to revert unwanted changes directly, but noting that it worked poorly for large changes.

Several participants discussed the \textbf{need to compare multiple prompts or AI-generated solutions side by side}, which is not supported by any existing tools. For example, P8 shared that when prompting multiple times, they hope to view different versions of the resulting code changes in a centralized way. P2 added that they prefer to have \participantquote{persistence of the chat} even when reverting the chat to a previous state, but the current tools \participantquote{will clear the history instead.} 
P6 said they would like a tool that allows them to map and revisit the \participantquote{different versions of the AI-generated answers as well as the different prompts corresponding to them.}

Together, these accounts highlight a mismatch between the fine-grained, exploratory nature of AI-assisted programming and the coarse and code-centric versioning mechanisms available in current tools. 
This observation motivates the design guideline:
\designguideline{DG1: Support the creation, comparison, and management of alternative solutions with minimal overhead.}

\subsubsection{Managing Prolonged AI interactions}
\label{subsubsec:challenge_long_interaction}
Participants described persistent challenges when working with AI assistants over long or complex coding sessions, finding they often \textbf{fail to understand user intention or follow prior instructions in long prompting sequences}.
P6 described their experience that \participantquote{by the third prompt, it already forgot what I asked in the first prompt}. To overcome the assistant's \participantquote{vanishing memory}, they have to \participantquote{append the specific keywords from the previous prompts} to help preserve key context across turns. P8 also found that AI tools struggle with more complex implementations and often fail to fully grasp what they intend it to do. To mitigate those issues, P7 suggested the need to break tasks into smaller, \participantquote{atomic} prompts.

Participants find it \textbf{challenging to provide effective instructions themselves and keep track of previous actions}, especially for complex tasks.
For example, P5 often blamed insufficient context in their own prompts for irrelevant or incoherent output of AI models.
P4 mentioned that they tend to \participantquote{forget what [they’ve] done} after using Claude Code for extended periods, making it difficult to recall prior changes, decisions, and partially implemented features.

Participants also expressed \textbf{a desire for AI assistants to have greater awareness of the exploration process}. For example, P4 suggested the need for the AI assistant to reference the history of what the developer has already tried, eliminating the manual effort of explaining their programming progress in the prompt. They also wished to have a system to directly preserve and manage long AI interaction states in order to reduce manual and cognitive efforts.

Together, these observations highlight how current tools struggle to support the management of long AI-assisted programming sessions, placing demands on both the AI assistant and the developer. 
These findings motivate the design guideline:
\designguideline{DG2: Support the review and management of evolving AI prompts and context across long interactions.}

\subsubsection{Determining Authorship and Provenance}
\label{subsubsec:challenge_provenance}
Participants described a wide range of perspectives on code authorship and responsibility in the context of AI-assisted programming. 
They had \textbf{different opinions on whether having the AI interaction history would impact their code review process}. 
P8 judged the AI-interaction trace could be \textbf{complicated for the reviewer to understand}, and if code they are reviewing \participantquote{works and it looks good, [I] don't really care if it came from an AI-generated source or how [the author] prompted it.} 
On the other hand, some stated that having the AI-interaction history helps \textbf{direct their attention} to potential errors when reviewing, so that they can determine \participantquoteattributed{which sections to look at more}{P1}.

On the individual level, participants indicated a more consistent view of the need for provenance tracking of AI-generated edits to help them conduct effective verification. They described their \textbf{challenges in differentiating AI-generated code from their own manually authored code} using existing tools. For example, P8 described finding it \participantquote{annoying} when uncommitted code was interleaved with new AI-generated content, making later separation difficult.
During extended coding sessions, P1 would like a tool that could show which \participantquote{sections of [the] code were [AI-generated] and [which] were handwritten,} as well as indicate \participantquote{which prompts were used to generate the code.} This provenance information would help them understand how the development process took place and revisit the reasoning behind specific changes.

Motivated by participants' need for verifying AI-generated edits in their programming sessions, we propose the design guideline:
\designguideline{DG3: Make the AI-generated code in the codebase explicit and easily identifiable for verification.}
\section{System Design Process}

To meet the developers' needs during AI-assisted programming, we undertook an iterative design process.

\subsection{Ideation and Initial System Design}

Informed by insights from prior literature on AI-assisted workflows, results from our preliminary interview study, as well as our own experiences using AI coding assistants, we set the overarching goal of the system as \textbf{to enable developers to better manage their iterative development process when performing AI-assisted programming}. To meet this overall goal, our initial design focuses on supporting developers in visualizing the relationships between code changes and AI interactions and allowing developers to flexibly switch between different exploration steps and paths.

We began by creating initial sketches and prototypes in Figma, going through multiple rounds of ideation and refinement within the research team. These iterations explored different ways of representing branching code paths, linking chat interactions to corresponding edits, and embedding these visualizations directly into the coding workspace.
We envisioned the tool as an integrated component operating alongside developers' AI assistant in their IDE. The tool would automatically capture the evolution of both code and AI conversations, by recording snapshots of both the state of the codebase and the chat history. Developers could then interact with this recorded history through a visual graph interface that supports branching, comparison, and restoration of prior development states without disrupting their ongoing workflow.

\subsection{Feedback to the Initial Prototype}
\label{subsec:feedback}
With the initial prototype, we conducted a formative user study with the same participants from the preliminary study (see Section~\ref{sec:preliminary_study}) to further understand user needs and iterate on the system design. 

First, their feedback validated the three underlying DGs and the initial features, as participants highlighted benefits related to transparency, exploration freedom, and the management of long interactions. P5 found the graph-based history would be \participantquote{very useful} for revisiting earlier prompts as conversations became large and noisy. Participants emphasized the value of making AI contributions explicit; P1 appreciated being able \participantquote{to see what was LLM-generated}. P2 described history management as helping developers \participantquote{not fear the code}, and P7 characterized the graph as giving \participantquote{credibility to the code} by exposing how it emerged from specific inputs. Participants also valued the ability to compare alternatives without losing context and therefore better support exploratory work.

At the same time, participants provided several suggestions that informed refinements to the final design. A primary concern involved cognitive load as the history graph grows.
P1 noted that large and complex graphs could become overwhelming, suggesting features such as automatically focusing on the most recent nodes to help maintain orientation. Additionally, P8 pointed out that creating nodes on every save could make the graph overly cluttered, motivating our later idea of explicit checkpoints triggered by users.

Participants also emphasized the importance of user control over the graph itself. For example, P1 expressed a desire to clean up the graph to improve readability over time. In response, we included checkpoint manipulation features allowing users to rename, merge, and delete nodes. Others highlighted the need to keep the system separate from traditional version control. P2 specifically noted that they would want the experimentation with AI assistance to not affect the project's actual repository state. Our final design reflects those considerations, making sure that the exploration history is local and does not interfere with traditional version control.

Finally, participants discussed how the system could better support context management across long time spans and during collaborative settings. 
P7 envisioned using the graph to return months later to the original chat history and reasoning behind a feature and continue building from that point.
Several participants also reflected on how enhanced context awareness could improve AI responses. For example, P8 suggested that the user should be able to provide the assistant with the existing development history as context to their prompts, an idea we implemented in the final version.

\begin{figure*}[t]
    \centering
    \includegraphics[width=.9\textwidth]{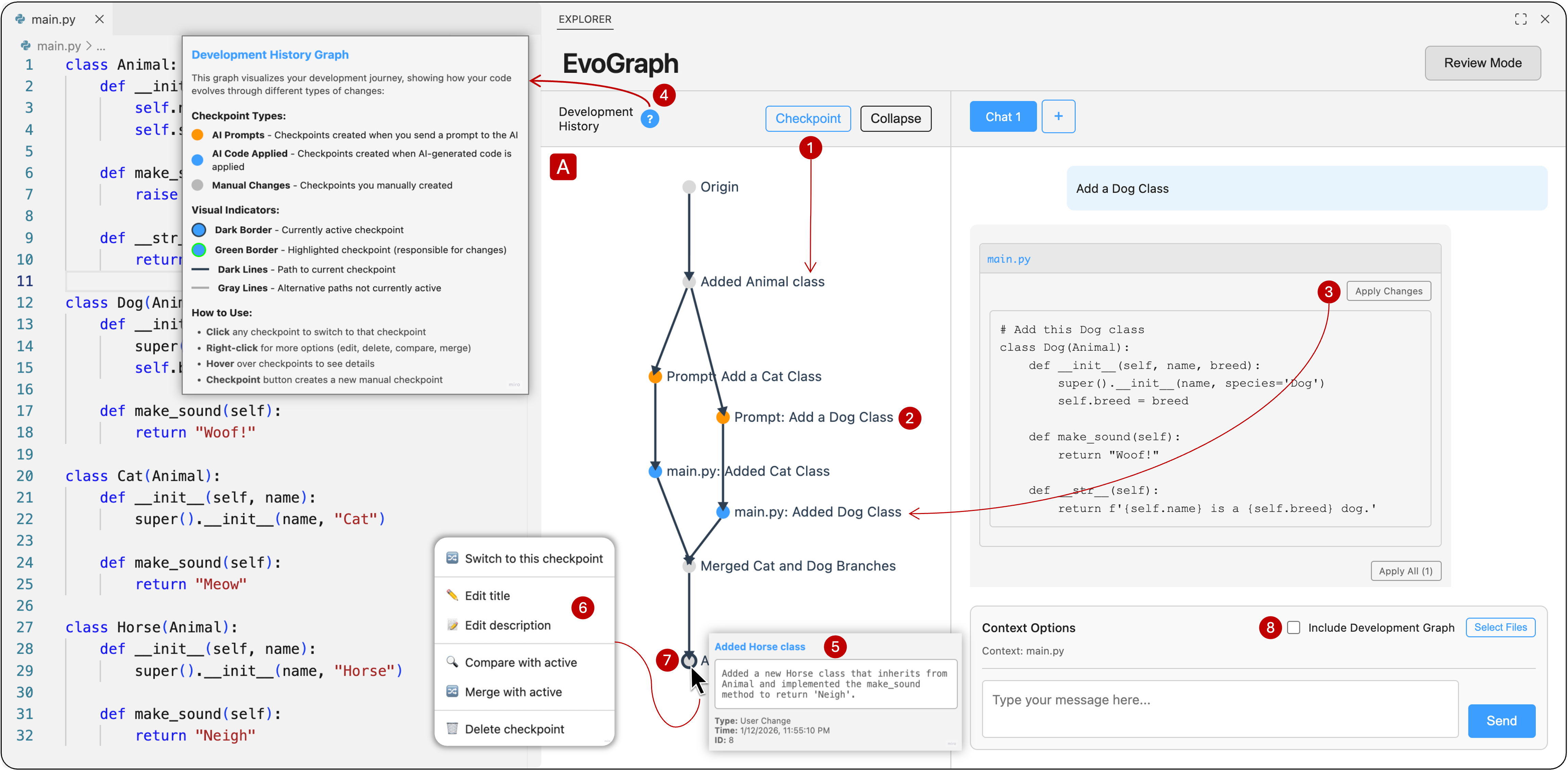}
    \caption{\toolname{} adds a development history graph and the surrounding features~\textcircled{A}. Nodes on the graph represent code and prompt checkpoints created either directly by the user~\textcircled{1} or automatically by prompting the AI assistant~\textcircled{2} and accepting AI-generated code~\textcircled{3}. Each checkpoint has an editable title and description~\textcircled{5} and~\textcircled{6}. The active checkpoint is indicated by a dark border~\textcircled{7}. Branches are created when the development flow is reversed to a previous checkpoint and continued along a different direction. The user can decide to include the entire branching history as context for future interactions with the AI~\textcircled{8}.}
    \label{fig:graphview}
    \Description{Screenshot of the EvoGraph interface integrated into a code editor. The interface is divided into three main regions. On the left, a Python file is displayed with class definitions (Animal, Dog, Cat, Horse). In the center, a vertical “Development History Graph” shows a sequence of checkpoints connected by branches. The graph begins at an “Origin” node and branches into different paths labeled with actions such as “Prompt: Add a Cat Class” and “Prompt: Add a Dog Class,” followed by nodes indicating applied code changes (e.g., “main.py: Added Cat Class”). The branches later merge into a single node labeled “Merged Cat and Dog Branches,” followed by an additional checkpoint (“Added Horse class”). Nodes are visually differentiated by color and icons to indicate prompt actions, AI-applied code, and manual changes. The currently active checkpoint is highlighted with a dark border. On the right, a panel shows AI-generated code for adding a Dog class, along with an “Apply Changes” button. Below this, a chat interface allows the user to send prompts, with an option to include the development graph as context.}
\end{figure*}

\section{System Description}
We developed a working version of \toolname{} as a VS Code extension that extends developers' familiar chat-based AI interaction to address breakdowns around versioning, context management, and provenance tracking identified in our preliminary study.

\subsection{User Interaction Design of \toolname{}}
The main functionalities of \toolname{} are achieved through (1) a \textit{development history graph} that persists both prior code and AI interaction states, (2) a \textit{checkpoint manipulation system} for managing the history, and (3) a \textit{review mode} for inline provenance tracking.

\subsubsection{The Development History Graph View}
The development history graph panel is displayed next to the familiar chat-based AI interaction panel (Figure~\ref{fig:graphview}~\textcircled{A}), to provide a historical view of the key information concerning the evolution of prompting and code changing actions [DG1, DG2]. 
The graph is built up with \textit{checkpoints}, each of which represents a snapshot of AI-assisted development at a particular point in time.

The system manages the following three types of checkpoints:
(1) \textbf{Manual change checkpoints} are created whenever the user clicks the \texttt{Checkpoint} button on top of the graph panel (Figure~\ref{fig:graphview}~\textcircled{1}). This provides flexibility for users to manually track code changes at their own pace. (2) \textbf{AI prompt checkpoints} are created automatically whenever the user sends a prompt to the AI assistant on the chat panel and receives a response from the AI (Figure~\ref{fig:graphview}~\textcircled{2}). This represents the AI interaction states and can occur regardless of whether the AI's response contains code. When the user edits a previous prompt in the chat history, an AI prompt checkpoint will be created as a new branch from the existing checkpoint.
(3) When the AI assistant suggests code changes, detected code blocks in the response will be rendered with an \texttt{Apply Changes} button on the chat panel (Figure~\ref{fig:graphview}~\textcircled{3}). Clicking on this button shows the applied changes as diff hunks in the editor and automatically creates a \textbf{AI code applied checkpoint}. 
The three types of checkpoints are differentiated by color coding, which is also explained to the user in a pop-up help panel (Figure~\ref{fig:graphview}~\textcircled{4}).

Each checkpoint is associated with a title and a description (Figure~\ref{fig:graphview}~\textcircled{5}); they are generated with the AI assistant according to checkpoint type: \textit{AI prompt checkpoints} use the user prompt and assistant response, while \textit{AI code applied} and \texttt{manual change} checkpoints summarize the code diff from the parent. The user can edit the title or description by right-clicking on a checkpoint and selecting the relevant action from the context menu (Figure~\ref{fig:graphview}~\textcircled{6}).

At any time, one checkpoint on the graph is in the \textit{active} state, represented by a dark border in the graph (Figure~\ref{fig:graphview}~\textcircled{7}), corresponding to the state of the codebase.
Left-clicking any other checkpoint switches the editor to the codebase state and AI assistant state associated with that checkpoint.
When a new checkpoint is created, either automatically or manually, it is directly linked to the previously \textit{active} checkpoint. If a checkpoint is created from a parent that already has a child, the new checkpoint forms a new branch without overwriting existing states [DG1].

When interacting with the AI Assistant, \toolname{} allows the user to include the Development History Graph as part of the prompt context through the panel's context options (Figure~\ref{fig:graphview}~\textcircled{8}). When this option is enabled, the assistant is provided with information about prior checkpoints, their relationships in the development graph, and the overall evolution of the codebase. 
This allows the assistant to generate responses that take into account development progress across multiple branches, rather than being limited to the current code state or active chat session [DG2].

\begin{figure*}[t]
    \centering
    \includegraphics[width=\textwidth]{system_screenshots/checkpoint_fixed.png}
    \caption{\toolname{} allows users to compare~\textcircled{1}, merge~\textcircled{2}, delete~\textcircled{3}, and switch~\textcircled{4} checkpoints.}
    \label{fig:checkpoint_manipulation}
    \Description{Screenshot demonstrating checkpoint manipulation features in EvoGraph. The figure shows multiple views of the development history graph and related interfaces. On the left, a vertical graph displays checkpoints such as “Origin,” “Added Animal class,” and branches for adding Dog and Cat classes, with nodes representing prompts and applied code changes. A context menu is shown next to a selected checkpoint, listing actions including switching to the checkpoint, editing title and description, comparing with the active checkpoint, merging with the active checkpoint, and deleting the checkpoint. Annotated callouts highlight four key interactions. (1) Compare: a side-by-side code comparison view showing differences between two checkpoints, with added and removed code highlighted in green and red. (2) Merge: two branches are combined into a new node in the graph; the code editor displays the changes to be accepted from the merge, and the chat history includes a new message summarizing the conversation being merged into the current context. (3) Delete: the graph with a checkpoint removed, resulting in a simplified branch structure. (4) Switch: selecting a checkpoint from the graph to restore the code and state associated with that checkpoint.}
\end{figure*}
\subsubsection{Checkpoint Manipulation}
The user can manipulate the checkpoints on the development history graph in the following ways, all through the context menu of a checkpoint.
First, by clicking on the \texttt{Compare with active} item on the context menu, a temporary diff file will be displayed as a VS Code editor tab showing differences in both the codebase and the associated AI interaction states between the active checkpoint and the currently selected checkpoint (Figure~\ref{fig:checkpoint_manipulation}~\textcircled{1}). This allows developers to inspect how code and interaction context evolved together [DG1, DG2].

Second, the \texttt{Merge with active} item will lead to the creation of a new checkpoint that combines the active checkpoint and the currently selected checkpoint. Both the code changes and their corresponding AI assistant chat histories will be merged (see Figure~\ref{fig:checkpoint_manipulation}~\textcircled{2}) [DG1, DG2].
For code, merging is performed by computing the diff between the checkpoints being merged and the most recent common ancestor shared by the two checkpoints. This diff is then provided to the AI assistant, which proposes a merged version of the code. 
For the AI assistant chat histories, the merge is handled by extending the chat history of the active checkpoint with a message summarizing the prior conversation from the other checkpoint and indicating that another development branch is merged.
By merging two existing checkpoints, the user creates a new manual checkpoint whose title and description are created automatically by combining the titles and summaries of the checkpoints being merged.

Further, the \texttt{Delete checkpoint} item on the context menu allows the user to delete the selected checkpoint. When a checkpoint is deleted, its children are reattached to the deleted checkpoint's parent (see Figure~\ref{fig:checkpoint_manipulation}~\textcircled{3}) [DG1, DG2]. Finally, the \texttt{Switch to this version} item (see Figure~\ref{fig:checkpoint_manipulation}~\textcircled{4}) functions the same as left-clicking on the checkpoint, to make it the active checkpoint [DG1].

\begin{figure}[t]
    \centering
    \includegraphics[width=\columnwidth]{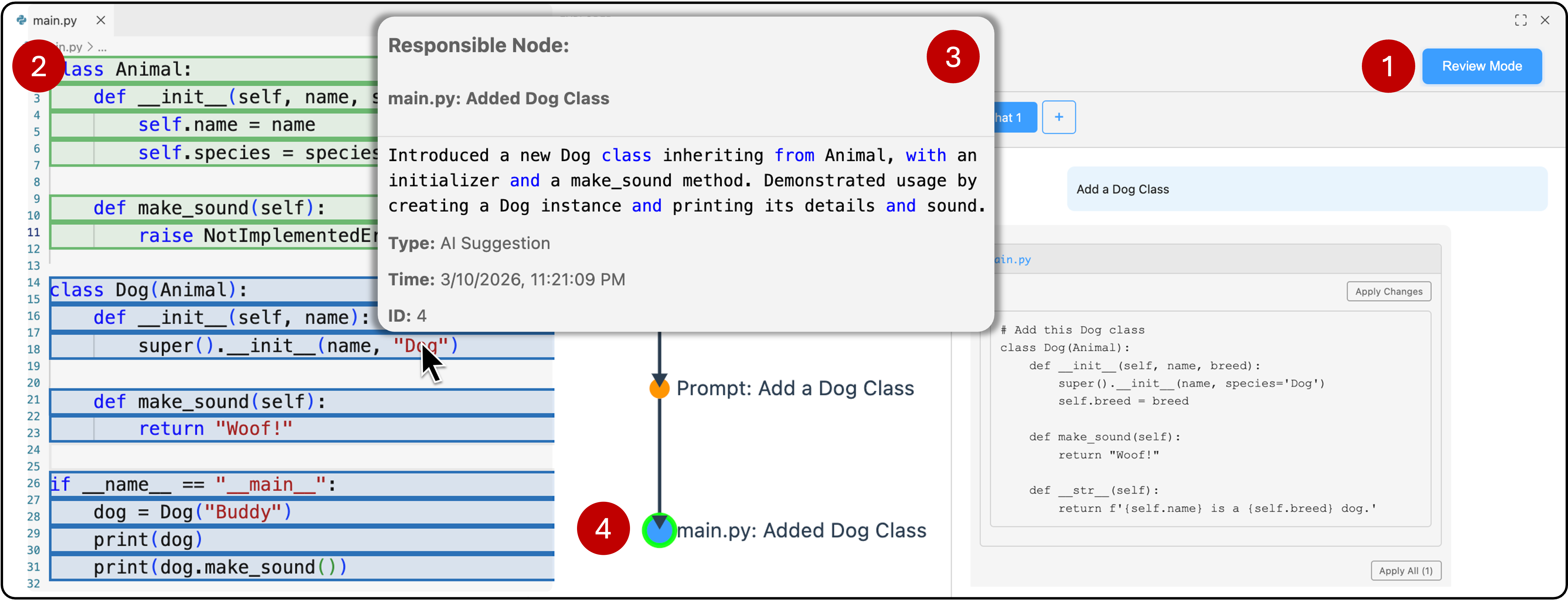}
    \caption{In review mode, \toolname{} highlights all code changes since the origin \circled{2}, indicating the author: green for human and blue for AI. Hovering over a highlighted code section displays information about the responsible checkpoint \circled{3} and highlights it in the development graph \circled{4}.}
    \label{fig:review_mode}
    \Description{Screenshot of EvoGraph in Review Mode within a code editor. The interface shows a Python file where code changes are highlighted to indicate their origin. Newly added code segments are outlined in different colors: blue indicates AI-generated changes, and green indicates human-authored changes. A “Review Mode” button is visible in the top-right corner. When the user hovers over a highlighted code section, a tooltip appears labeled “Responsible Node,” displaying metadata about the change, including the file name (“main.py: Added Dog Class”), a textual description of the modification, its type (AI suggestion), timestamp, and an identifier. Simultaneously, the corresponding checkpoint node in the development history graph is highlighted, linking the code segment to its origin in the graph.}
\end{figure}

\subsubsection{Review Mode}
The Review Mode can be activated on the panel (Figure~\ref{fig:review_mode}~\textcircled{1}), which provides inline provenance information directly in the code editor [DG2, DG3]. Once activated, all code changes since the origin checkpoint will be highlighted, distinguishing between AI-generated and human-edited changes [DG3]. The system computes line-by-line diffs between the active checkpoint and the origin checkpoint. Lines added or modified by AI-generated checkpoints are highlighted in blue, while human-edited lines are highlighted in green. (See Figure~\ref{fig:review_mode}~\textcircled{2}.)

Hovering over a highlighted line displays a tooltip containing information about the responsible checkpoint, including its title, description, type, and timestamp (Figure~\ref{fig:review_mode}~\textcircled{3}) [DG2]. At the same time, the corresponding checkpoint is highlighted directly in the Development History panel using an animated pulsating green ring border effect (see Figure~\ref{fig:review_mode}~\textcircled{4}), visually linking code changes to their provenance in the interaction history [DG3].

\subsection{Implementation}
The tool is implemented as a VS Code extension using TypeScript, with two main webview-based panels that communicate with the extension host through message passing. The AI Assistant Panel integrates with OpenAI's GPT-4o model for chat responses and code generation, while the Development History Panel renders an interactive graph visualization of the development history. State management is handled through VS Code's Extension Context API, which persists development history, chat sessions, and file snapshots across extension reloads. The diff preview system uses VS Code's text editor decorations and CodeLens providers, and the Review Mode leverages hover providers.

OpenAI's GPT-4o model is used throughout the system for generating chat responses, applying code edits, creating checkpoint titles and descriptions, and merging checkpoints. The model is fed with the diff between a checkpoint and its parent to generate its title and description. For applying code edits, the model receives the code block to be applied and the current file and is prompted to return the updated file with the changes applied. For merging, it is provided with the diffs of both checkpoints relative to their most recent common ancestor and is asked to produce the merged file.

\section{User Study}

The value of \toolname{} lies in supporting the process of exploration, sense-making, and error recovery, rather than optimizing final outputs. Therefore, our user study focuses on understanding how \toolname{} shapes participants' experiences in exploring emerging ideas and managing evolving code while working with AI assistants.

\subsection{Methods}

\subsubsection{Participants}

We conducted a task-based, within-subjects user study with
20 participants who were experienced in web development and familiar with AI-assisted programming tools (three aged 18–20, eleven 20–24, and six 25–29; 9 female, 11 male). They were recruited through computer science–related channels and personal networks.
Participants reported frequent use of AI programming assistants, with 11 using them daily and nine a few times per week. Their software development experience ranged from one to ten years (six with 1–3 years, seven with 3–5 years, and seven with 5–10 years). 
All sessions were conducted remotely, and participants received CA\$45 as a token of appreciation.

\subsubsection{Procedure}
Each session lasted approximately 90 minutes and consisted of two programming tasks. Participants completed the two tasks using either the full \toolname{} or a Control interface without graph or review features. The Control interface resembled existing AI programming assistants such as GitHub Copilot or Cursor, providing a chat-based interaction panel, as well as a reversion mechanism through editing previous messages, which resets the conversation state and associated code changes to that point. To reduce the chance of demand characteristics~\cite{demandChars}, the tool name in the header was replaced by \textit{StudyTool} and the two conditions were simply referred to as \textit{v1} and \textit{v2}. Before starting each task, participants received a short tutorial to familiarize themselves with the version of the tool they would be using. 

The order of task-tool combinations was counterbalanced using Graeco-Latin square design~\cite{montgomeryDesign}. The programming tasks were intentionally open-ended, giving participants freedom to explore alternative solutions while working toward concrete feature implementations with clearly defined goals. For each task, participants were asked to clone a provided starter repository and were given 25 minutes to add features to enhance the application. In one task, participants were asked to enhance a basic weather dashboard application, and in the other, a task management dashboard. Participants were free to decide which features to implement and how to implement them. For both tasks, they were given access to the assigned tool and Git, as well as their browser to view the application. After completing each task, participants were asked to write a short report describing the features they implemented and how they worked, during which they could revisit the code and their prior interactions with the AI assistant.

Following each task, participants completed a NASA-TLX questionnaire to assess perceived workload. They then responded to a set of Likert-scale questions assessing their experience with (1) exploring alternative solutions (four questions, on the ease of reverting changes, as well as trying, comparing, and combining alternative solutions), (2) prompt management (three questions, on the ease of reviewing chat history, refining prompts, and providing context to the assistant), and (3) provenance tracking (three questions, on the ease of tracing the origin of changes, reviewing the progression of the codebase, and overall awareness of changes made). The questionnaires are provided in Appendix~\ref{sec:appendix}. Finally, participants were asked to compare their experiences in both conditions and discuss perceived benefits and drawbacks in each.

\subsubsection{Analysis}
All sessions were audio- and screen-recorded. We analyzed qualitative data using inductive thematic analysis~\cite{braun_thematic_2021}. The first author conducted open coding of interview transcripts and task sessions, using screen recordings to contextualize participants' actions and interactions during AI-assisted programming. Codes were iteratively refined and grouped using affinity diagramming to identify higher-level themes and clusters. These themes were then discussed and refined collaboratively by all three authors through multiple rounds of discussion until consensus was reached and we were satisfied with the coherence and prominence of the themes.

We analyzed quantitative data using linear mixed effects models in R, examining fixed effects of Tool, Task, and Order (i.e., independent variables) and all interaction effects between them, while controlling for the random effect of Participant. For each dependent variable, we first fit a full model with all interactions and the random effect. We then tested whether the random effect of Participant was necessary by comparing the linear mixed model to a linear model without the random effect using likelihood ratio tests. When the random effect was not significant, we simplified to a linear model and further tested whether interaction terms could be removed. All $\alpha$ levels were set at $0.05$. P-values for main and interaction effects were calculated using the Kenward-Roger method for degrees of freedom and reported from ANOVA tables with \texttt{lmerTest}~\cite{lmerTest}. Post-hoc tests were conducted via estimated marginal means (\texttt{emmeans}) with Bonferroni correction. 
We visually inspected the normality of model residuals (via Q-Q plots) and the assumption of homoscedasticity (via spread-level plots of absolute residuals against fitted values).

\begin{figure*}[t]
    \centering
    \begin{subfigure}[t]{0.44\textwidth}
        \centering
        \includegraphics[width=\linewidth]{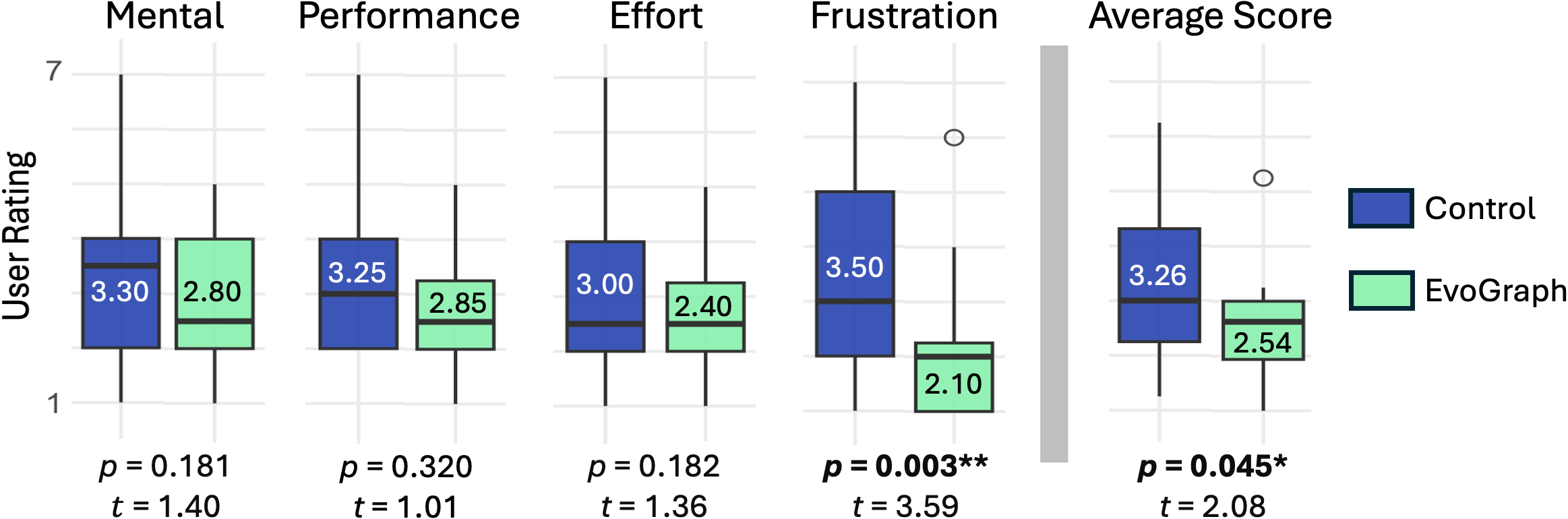}
        \caption{Cognitive load (NASA-TLX) dimensions}
        \label{fig:nasa_tlx_plots}
        \Description{Cognitive load (NASA-TLX) dimensions. EvoGraph shows lower scores than Control across all measures: Mental (3.30 vs 2.80, p = 0.181), Performance (3.25 vs 2.85, p = 0.320), Effort (3.00 vs 2.40, p = 0.182), and Frustration (3.50 vs 2.10, p = 0.003). The average score is also lower for EvoGraph (3.26 vs 2.54, p = 0.045), indicating reduced overall cognitive load.}
    \end{subfigure}
    \hfill
    \begin{subfigure}[t]{0.44\textwidth}
        \centering
        \includegraphics[width=\linewidth]{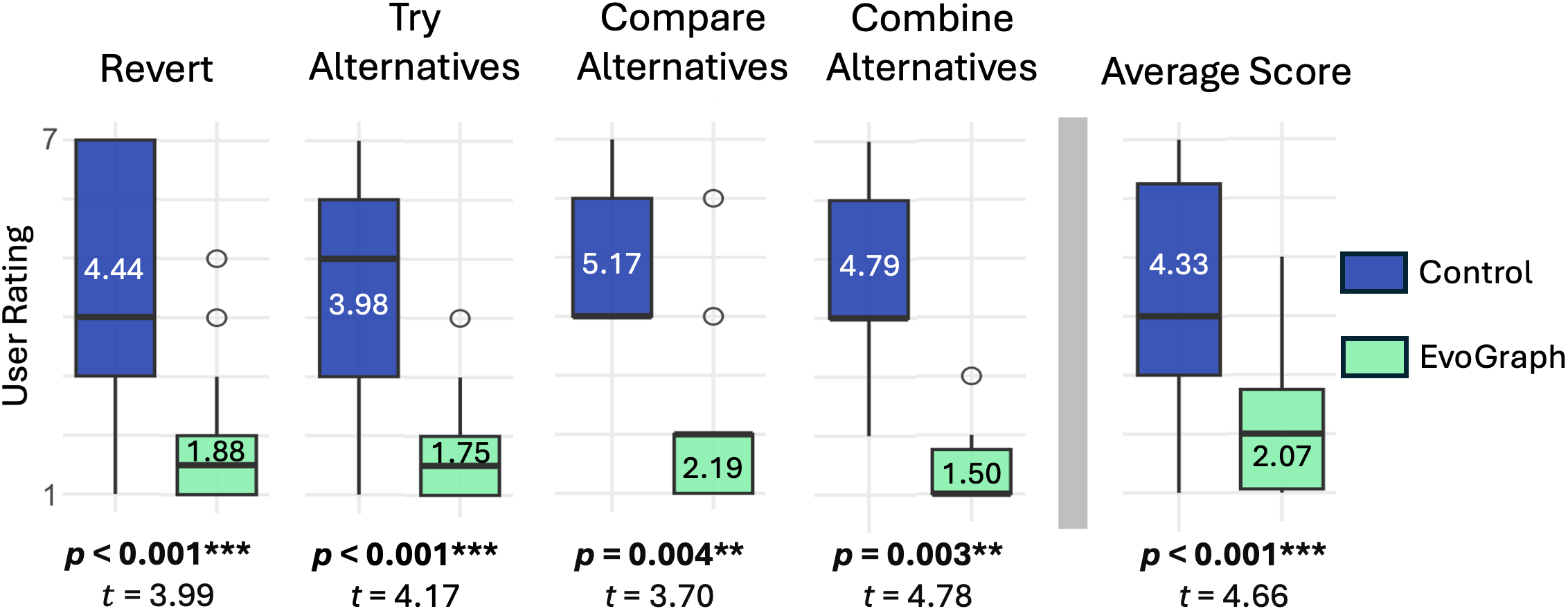}
        \caption{Alternative exploration dimensions}
        \label{fig:exploration_plots}
        \Description{Alternative exploration dimensions. EvoGraph consistently achieves lower scores than Control: Revert (4.44 vs 1.88, p < 0.001), Try alternatives (3.98 vs 1.75, p < 0.001), Compare alternatives (5.17 vs 2.19, p = 0.004), and Combine alternatives (4.79 vs 1.50, p = 0.003). The average score is substantially lower for EvoGraph (4.33 vs 2.07, p < 0.001), indicating improved support for exploring alternatives.}
    \end{subfigure}

    \vspace{0.5em}

    \begin{subfigure}[t]{0.44\textwidth}
        \centering
        \includegraphics[width=\linewidth]{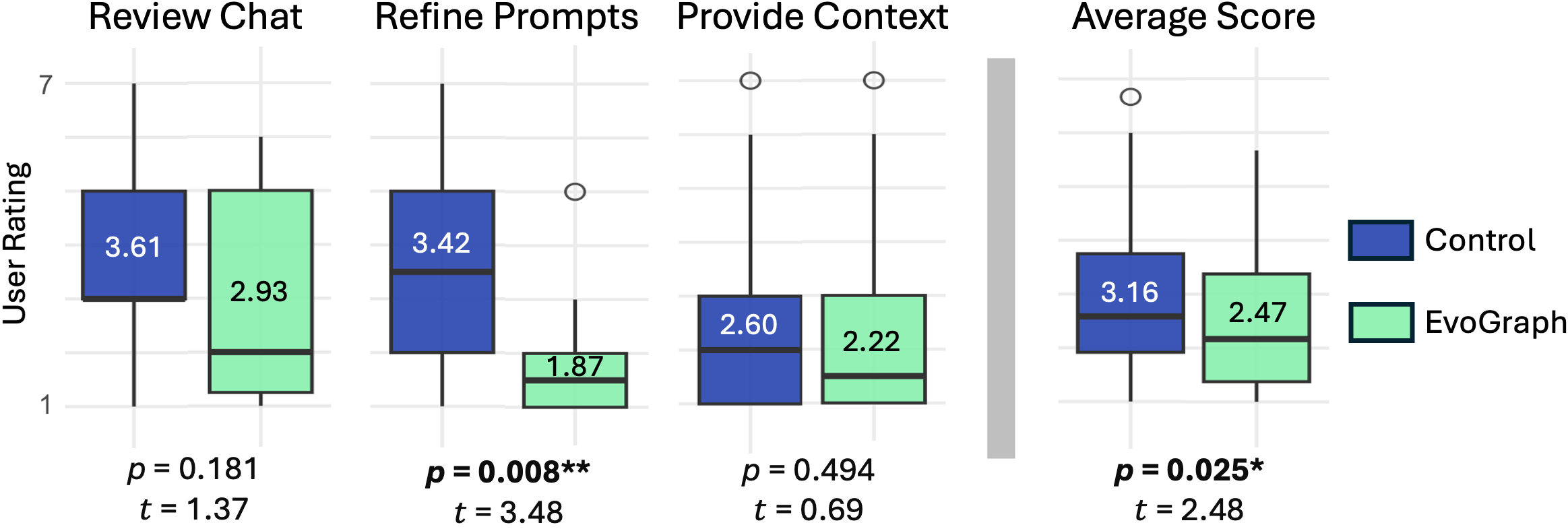}
        \caption{Context management dimensions}
        \label{fig:context_plots}
        \Description{Context management dimensions. EvoGraph shows lower scores for Review chat (3.61 vs 2.93, p = 0.181), Refine prompts (3.42 vs 1.87, p = 0.008), and Provide context (2.60 vs 2.22, p = 0.494). The overall average is also lower for EvoGraph (3.16 vs 2.47, p = 0.025), indicating improved context management.}
    \end{subfigure}
    \hfill
    \begin{subfigure}[t]{0.44\textwidth}
        \centering
        \includegraphics[width=\linewidth]{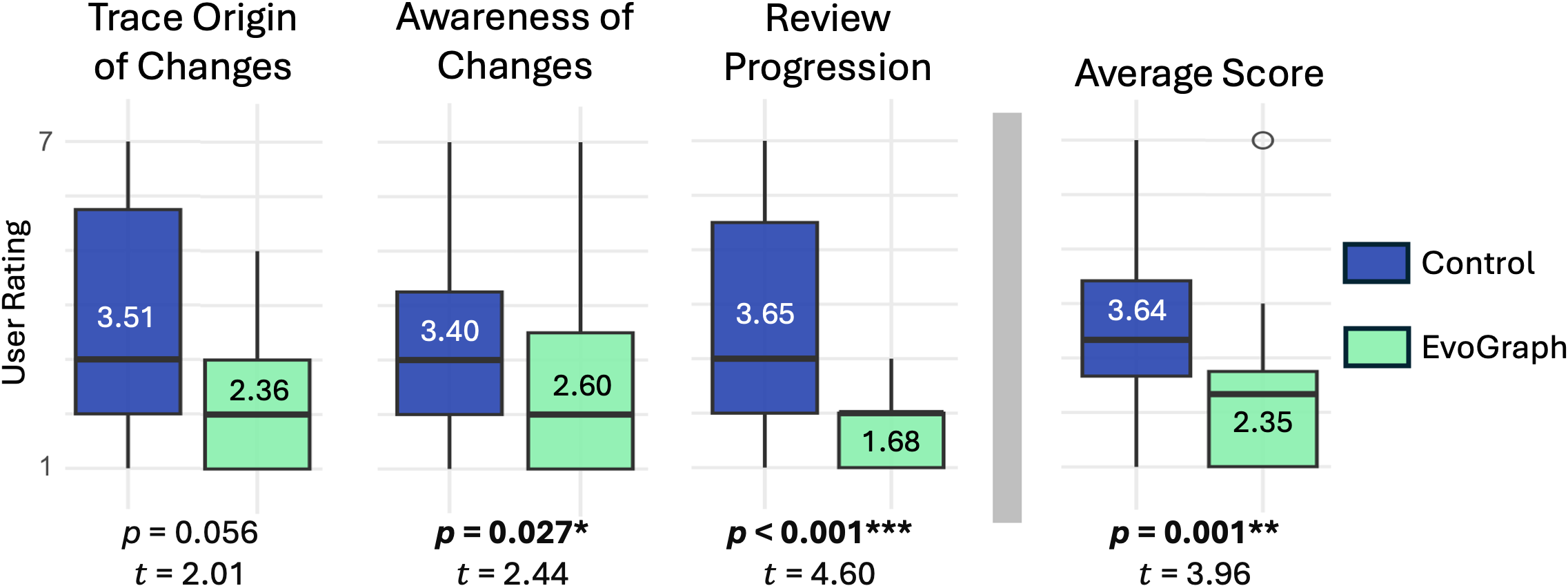}
        \caption{Provenance tracing dimensions}
        \label{fig:provenance_plots}
        \Description{Provenance tracing dimensions. EvoGraph shows lower scores across all measures: Trace origin of changes (3.51 vs 2.36, p = 0.056), Awareness of changes (3.40 vs 2.60, p = 0.027), and Review progression (3.65 vs 1.68, p < 0.001). The average score is also lower for EvoGraph (3.64 vs 2.35, p = 0.001), indicating improved provenance tracking.}
    \end{subfigure}
    
    \caption{Users' ratings across four dimensions. Low values indicate low cognitive load or difficulty and are preferred in all dimensions. Boxplots display estimated marginal means. Estimated marginal means, t-statistics, and p-values are based on post-hoc tests with Bonferroni correction reported from the \texttt{emmeans} R package. (***p < 0.001, **p < 0.01, *p < 0.05)}
    \Description{Four sets of boxplots comparing user ratings between two tools: Control (blue) and EvoGraph (green). Lower values indicate lower cognitive load or difficulty and are preferred. Each plot shows distributions along a 1–7 scale, with estimated marginal means labeled inside the boxes. Across most dimensions, EvoGraph shows lower (better) scores than Control, with statistically significant differences indicated by p-values and t-statistics.}
    \label{fig:quantitative_results}
\end{figure*}

\subsection{Quantitative Results}
Figures~\ref{fig:nasa_tlx_plots}--\ref{fig:provenance_plots} summarize the statistical test results on the participants' ratings between the two conditions. Compared to the Control, the participants found that \toolname{} significantly (1) reduced cognitive load, especially reducing frustration, (2) allowed for easier exploration of alternatives, supporting various exploratory activities, (3) facilitated prompts and context management, particularly for refining prompts, and (4) aided tracing of changes during AI-assisted programming, fostering awareness and review.

We also observed an ordering effect across three overall scales (\textit{Alternative Exploration}, \textit{Context Management}, and \textit{Provenance Tracing}), where the contrast between \toolname{} and the Control was larger during the participants' second task. Among the specific questions in which \toolname{} exhibited a better outcome, the ordering effect was observed on two items, \textit{try alternatives} and \textit{review progression}. We speculate that this could be due to a variety of factors, such as a lack of grounding for the subjective scores during the first task, a learning effect, or demand characteristics.

\subsection{Qualitative Results}
We identified several themes on how participants experienced \toolname{} in action, how they perceived it could be further improved.

\subsubsection{Effects of \toolname{} on Participants' AI-Assisted Programming Workflows}
We found that \toolname{} had a positive impact on the AI-assisted coding process, with participants overwhelmingly preferring it over the Control. 
Participants also expressed strong interest in adopting the tool in practice, with several asking whether \toolname{} would be released publicly and requesting permission to continue using it for personal projects. We observe several emerging benefits of \toolname{}, either explicitly stated by the participants or reflected in their usage during the programming tasks.

\textbf{Enabling Safe Exploration and Better Leverage of Interaction History.}
Participants using \toolname{} expressed greater confidence in experimenting with different ideas, knowing that prior states could be preserved and revisited. As P2 explained, \participantquote{You can switch context pretty fast without having to discard what you've done before. I think that's the main advantage, because you can always discard stuff in normal code editors which have AI assistance, but then to go back to the previous data is not as easy.}
Without the constant need to safeguard existing code, participants perceived a reduced cognitive load during the AI-assisted programming. P10 elaborated on this benefit: \participantquote{if I do everything in their own branches, I know which feature influences which part. [...] But if I do it linearly and something goes wrong, it may be from something I did at the beginning and it's hard to go back.}

Participants also discussed how including the development graph as context for the AI assistant supported exploration, observing AI responses more relevant to the entire exploration history, as opposed to being limited to the current selected path.
For example, after exploring multiple implementation attempts for a feature across different branches, P8 included the development graph as part of the prompt context to \participantquote{ask what went wrong previously so [they] can learn from this mistake.}
Some participants utilized this functionality to ensure the assistant would not lose focus on the overall goal of the project. As P17 explained, \participantquote{[if] you merely rely on the chat history, then it's really easy to get lost, probably for both human and AI. And I think by giving the graph it will give less distraction for both human and AI, so the AI always knows what you're really doing and [...] always focuses on the right flow, the main flow, the main stream.}

In contrast, several participants noted that, when using the Control tool, fear of breaking or losing working code made them less likely to continue exploring alternative solutions. P4 explained such hesitation, even though they were \participantquote{still not satisfied from the result}, they were \participantquote{not sure [the AI assistant] could make a better design.} Participants using the Control tool also described difficulty managing partially completed work. As P4 explained, if they did not reach a working version, they could not leave the work unfinished and return to it later without losing progress; in their current workflow using Claude Code, this limitation sometimes resulted in them needing to \participantquote{remove all the ideas from three hours or four hours [of work].}

\textbf{Improving Efficiency for Iterating on Changes.}
Participants perceived \toolname{} allowed them to more efficiently create, compare, and refine alternatives. By switching between nodes of the development graph, P1 explained that they tried and compared \participantquote{five different solutions and then continue down the path of the one I want.} Similarly, when deciding on a visualization style for the weather chart, P13 created three alternatives in parallel, switched between them to compare the rendered visualizations in the browser, and continued development from the version they preferred.

When encountering incorrect code changes suggested by the AI assistant, participants found \toolname{} easier for them to revert changes than the Control, because they could just \participantquoteattributed{use the GUI to click [the node] and go back} {P8}.
Especially as AI-generated changes accumulated across several prompts, the graph-based reversion helped the participants quickly identify where the implementation began to deviate from their intent.
As P20 explained, \toolname{} allowed them to \participantquote{go straight down that path and just keep trying to hammer away at it}, while still being able to \participantquote{always just revert back to the first node}. P20 noted that this was especially important because when incorrect changes compound, and without such a mechanism, they would  \participantquote{be completely stuck} with failing code.

In contrast, participants found the Control tool's linear and one-directional reversion mechanism destructive (typically seen in the current programming-assistant tools like Copilot and Cursor)---it was difficult to recover once participants reverted beyond their intent. P11 described after a multi-prompt implementation effort went awry, they scrolled through the chat history and reverted too far, resulting in the loss of working code.

\textbf{Supporting Comprehension and Encouraging Reflection.}
Participants reported that reviewing the progression of the codebase was considerably easier with \toolname{} than with the Control tool. 
For example, P18 explained that the high-level overview enabled them to \participantquote{quickly see, on the graph, generalized headings of what each stage of my development was} without needing \participantquote{to manually go and sift through the chat.}
They also found that the information provided in the graph supported reasoning about how previous changes interacted. As P17 described, the graph provided \participantquote{more access points} to relevant information; this enabled a \participantquote{more versatile method to learn more about what you're doing and what would be the next thing to do.} While still avoiding overload, as they could access \participantquote{all the extra information without being distracted or getting lost}.

Participants mentioned that \toolname{} increased their awareness of the changes made to the codebase, encouraging them to critically reflect on the quality of the solution.
As P1 explained, \participantquote{ [\toolname{}] was so much better for a developer actually understanding what the LLM does. Nowadays, like a lot of programmers, we like to believe the LLM blindly and accept whatever code is being generated or added in. But I think [\toolname{}] is very useful at tracing the history of everything and really [allow us to have] a deeper understanding of the code being written.}

When using the Control tool, however, participants had to review the history by scrolling through the entire chat backwards; they found these long conversation histories with a large amount of text tough to navigate and make sense of. As P8 explained, \participantquote{when I wanted to go back and see the start of the response, I sometimes would get lost if I'm scrolling too fast}.

\subsubsection{Feedback for Future Improvements}
Several participants discussed the potential to extend \toolname{} to directly support collaborative settings. 
For example, P12 explained that having access to AI-generated summaries or chat histories alongside code changes during code review would help reviewers understand the process that led to a given change. In addition, participants noted that sharing a development graph within a team could make parallel work more transparent. P10 described that the node-based structure would make it easier for team members to understand what others had worked on, what had changed, and how different features were implemented in parallel. They emphasized that this visibility could be particularly useful when multiple developers are working on the same project and need to coordinate changes or resolve conflicts.

Participants also identified several areas for improvement related to usability and flexibility of the interface. 
Some participants noted that the inclusion of the history graph introduced additional visual complexity, as more information on the screen now needs to be managed alongside the code and chat interface. However, this was generally framed as a tradeoff of the added functionality rather than a fundamental usability concern.
Participants also highlighted interaction challenges tied to the \toolname{} supported workflow. For example, P19 reported forgetting to manually create a checkpoint after making changes, which resulted in their manual edits becoming intertwined with subsequent AI-generated changes.

Finally, participants voiced the need for greater control over the graph. They desired features that allow them to reorganize the graph layout (e.g., dragging nodes), as well as simplify it over time by combining or ``squashing'' multiple nodes. These suggestions point to the importance of supporting long-term maintainability of the graph as it grows in size and complexity.

\section{Discussion}
Our study reveals that \toolname{} significantly enhances the developer experience during AI-assisted programming---it boosted our participants' confidence in exploration and allowed them to efficiently iterate on and review their changes. Below, we first reflect on the design guidelines that informed \toolname{}. We then discuss broader implications of our work, especially in the constantly evolving context of AI-assisted programming.

\subsection{Reflections on the Design Guidelines}

\subsubsection{Reflections on DG1: Managing Alternative Solutions}

This guideline aims to address the mismatch between the fine-grained, exploratory nature of AI-assisted programming and the coarse versioning mechanisms available in current tools. \toolname{} follows this guideline by offering a development history graph that allows users to navigate, compare, and manipulate different solution paths, and the final user study demonstrated its effectiveness: participants rated the difficulty of performing various alternative exploration actions with \toolname{} significantly lower than with the control tool and reported increased confidence.
These findings demonstrate the importance of integrating flexible, micro-level versioning mechanisms directly into AI interactions. The versioning features provided by \toolname{} are still preliminary. Future tools could explore richer mechanisms integrating automated reasoning and manual input, such as semantic snapshots~\cite{DBLP:conf/kbse/Jiang19} and social features like forking~\cite{DBLP:conf/icse/ZhouVK20a}. Balancing the richness and flexibility in versioning support with the dynamic and fluid nature of AI programming remains key.

\subsubsection{Reflections on DG2: Managing Evolving Prompts and Context}
\toolname{} applies this guideline through a unified mechanism to manage code and prompting context together, enabling users to seamlessly track how prompts and code co-evolve over time and leverage the evolution path as prompting context. Participants considered that these features supported comprehension of programming progress and allowed focused exploration during AI-assisted programming. However, we did not observe a significant difference in their ratings on some scales regarding context management, such as the ease of reviewing prompting history and supplying development context to the AI. Together, these findings indicate the importance of having more structured, navigable representations of interaction histories in AI-assisted programming. Ideas from interactive prompt management tools (e.g., ChainForge for prompt comparison and chaining~\cite{Arawjo2024ChainForge} and OnGoal for goal tracking~\cite{Coscia2025OnGoal}) can be leveraged. Additionally, future tools could explore mechanisms that give users greater control over which programming exploration paths to include as prompting context to allow more flexible and intentional use of interaction history.

\subsubsection{Reflections on DG3: Tracing Provenance of Code}
To meet this guideline, we introduced inline provenance tracking through the Review Mode, which annotates code with both its author and the interaction from which it originated.
Participants made limited use of this feature during the study, likely due to the short task duration and the fact that most of the code was AI-generated. Yet, they commented on its potential value, particularly for supporting comprehension and encouraging critical reflection.
Moreover, several participants highlighted the potential of provenance tracking in collaborative settings, noting that combining provenance information with the development history graph could improve team awareness.
These findings suggest that provenance tracking needs to go beyond the AI-human dichotomy and embed richer information, such as requirements, design rationale, and possible tradeoffs; some of them may be extracted from the prompting history or the software documentation. Software engineering techniques for traceability~\cite{guo2025natural, DBLP:conf/icse/0004CC17} can be leveraged to identify relevant information sources and inform the design of more structured representations.

\subsection{Broader Implications for AI-Assisted Programming}
AI is continuously transforming programming work. 
Recent advances in agentic systems further extend AI's involvement by allowing these systems to autonomously plan and execute multi-step programming tasks~\cite{agenticCoding}. While these agents can accelerate development, they make it even harder for developers to maintain awareness and control over intermediate steps. Tools like \toolname{} could complement these systems by providing a structured representation of the agent's actions and corresponding project status, allowing developers to inspect, rewind, and intervene at any point in the workflow. Rather than restarting when an agent goes off track, developers could steer it back by revisiting and modifying earlier steps, enabling more control and transparency.

As AI becomes deeply embedded in software engineering, concerns are raised around authorship, traceability, and accountability~\cite{AccountabilityInCodeReview}. Our work can inform future efforts to address these challenges by expanding the scope of provenance beyond code changes, supporting developers in understanding how decisions are mediated by AI over time, thereby providing the opportunity to surface and correct systematic issues. This is especially prominent for domains that are highly regulated, such as those subject to Canada’s Directive on Automated Decision-Making~\cite{tbs2025directive} and the EU AI Act~\cite{eu2024aiact}, which emphasize transparency and traceability. By extending these principles to the actual process of AI-assisted programming, our approach contributes toward more accountable, maintainable, and trustworthy software development practices.

Finally, while collaboration is innate in software development, current AI coding assistants are built around a dyadic human–AI model and often overlook AI-assisted collaboration between humans. Rather than focusing on a collaborative chat interface (e.g.,~\cite{mao2024multiuserchatassistantmuca}), our tool opens new opportunities for collaboration in AI-assisted development by making individual interaction histories explicit, shareable, and interpretable across team members. These records can be embedded in repositories to allow developers to revisit prior prompts, understand the rationale behind design decisions, and build on each other’s work, supporting shared understanding and common ground~\cite{clark1991grounding, Olson01092000}.
However, this approach may raise certain sociotechnical issues that require future investigation. For example, increased transparency may introduce tensions, as developers may be hesitant to fully expose their reliance on AI due to concerns about perceived competence or judgment from peers.

\subsection{Limitations}

First, our study focused on a web development task, which may limit the generalizability of our findings to other domains or development contexts. Second, as our primary goal was to understand how \toolname{} supports developers’ workflows rather than its impact on code correctness or performance, we did not evaluate the quality of the code produced by the participants. Moreover, the tasks were open-ended and low-stakes, which may have led participants to place less emphasis on code quality than they would in real-world, production settings. Finally, our quantitative measures relied on subjective self-reported ratings, which may be influenced by individual perception and bias. Future work could explore more diverse tasks, investigate impacts on code quality, and evaluate the tool in more realistic development environments.

\section{Conclusion}

Our work presented \toolname{}, a system that incorporates a graph-based representation of local AI-assisted development history, enabling developers to easily capture, navigate, and compare alternative solutions while maintaining links between prompts and code.
Through a task-based user study, we found \toolname{} allows safe exploration and efficient iteration of solutions, improves the management of long interaction sequences, and increases awareness of AI-generated changes, while reducing cognitive load.
Grounded in our design guidelines, our findings have broader implications for future improvements to AI programming assistants in their constantly evolving landscape.

\balance
\bibliography{reference}

@inproceedings{NamMHVM24_LLM_for_Code_Understanding,
  author       = {Daye Nam and
                  Andrew Macvean and
                  Vincent J. Hellendoorn and
                  Bogdan Vasilescu and
                  Brad A. Myers},
  title        = {Using an {LLM} to Help With Code Understanding},
  booktitle    = {Proceedings of the 46th {IEEE/ACM} International Conference on Software
                  Engineering, {ICSE} 2024, Lisbon, Portugal, April 14-20, 2024},
  pages        = {97:1--97:13},
  publisher    = {{ACM}},
  year         = {2024},
  doi          = {10.1145/3597503.3639187},
}

@article{Barke_Grounded_Copilot,
author = {Barke, Shraddha and James, Michael B. and Polikarpova, Nadia},
title = {Grounded Copilot: How Programmers Interact with Code-Generating Models},
year = {2023},
issue_date = {April 2023},
publisher = {Association for Computing Machinery},
address = {New York, NY, USA},
volume = {7},
number = {OOPSLA1},
url = {https://doi.org/10.1145/3586030},
doi = {10.1145/3586030},
journal = {Proc. ACM Program. Lang.},
month = apr,
articleno = {78},
numpages = {27}
}

@inproceedings{Spellburst,
author = {Angert, Tyler and Suzara, Miroslav and Han, Jenny and Pondoc, Christopher and Subramonyam, Hariharan},
title = {Spellburst: A Node-based Interface for Exploratory Creative Coding with Natural Language Prompts},
year = {2023},
isbn = {9798400701320},
publisher = {Association for Computing Machinery},
address = {New York, NY, USA},
url = {https://doi.org/10.1145/3586183.3606719},
doi = {10.1145/3586183.3606719},
booktitle = {Proceedings of the 36th Annual ACM Symposium on User Interface Software and Technology},
articleno = {100},
numpages = {22},
location = {San Francisco, CA, USA},
series = {UIST '23}
}

@inproceedings{Pu_Assistance_or_Disruption,
author = {Pu, Kevin and Lazaro, Daniel and Arawjo, Ian and Xia, Haijun and Xiao, Ziang and Grossman, Tovi and Chen, Yan},
title = {Assistance or Disruption? Exploring and Evaluating the Design and Trade-offs of Proactive AI Programming Support},
year = {2025},
isbn = {9798400713941},
publisher = {Association for Computing Machinery},
address = {New York, NY, USA},
url = {https://doi.org/10.1145/3706598.3713357},
doi = {10.1145/3706598.3713357},
booktitle = {Proceedings of the 2025 CHI Conference on Human Factors in Computing Systems},
articleno = {152},
numpages = {21},
location = {
},
series = {CHI '25}
}

@inproceedings{liang2023largescalesurveyusabilityai,
author = {Liang, Jenny T. and Yang, Chenyang and Myers, Brad A.},
title = {A Large-Scale Survey on the Usability of AI Programming Assistants: Successes and Challenges},
year = {2024},
isbn = {9798400702174},
publisher = {Association for Computing Machinery},
address = {New York, NY, USA},
url = {https://doi-org.proxy3.library.mcgill.ca/10.1145/3597503.3608128},
doi = {10.1145/3597503.3608128},
booktitle = {Proceedings of the IEEE/ACM 46th International Conference on Software Engineering},
articleno = {52},
numpages = {13},
location = {Lisbon, Portugal},
series = {ICSE '24}
}

@misc{Cursor2025,
  title        = {{Cursor}},
  howpublished = {\url{https://cursor.com/home}},
  note         = {Accessed: 2025-12-10},
  year         = 2025
}

@misc{Copilot2025,
  title        = {{GitHub Copilot}},
  howpublished = {\url{https://github.com/features/copilot}},
  note         = {Accessed: 2025-12-10},
  year         = 2025
}

@misc{ClaudeCode2025,
  title        = {{Claude Code}},
  howpublished = {\url{https://claude.com/product/claude-code}},
  note         = {Accessed: 2025-12-10},
  year         = 2025
}

@article{SERGEYUK_AI_assistants_in_practice,
title = {Using AI-based coding assistants in practice: State of affairs, perceptions, and ways forward},
journal = {Information and Software Technology},
volume = {178},
pages = {107610},
year = {2025},
issn = {0950-5849},
doi = {https://doi.org/10.1016/j.infsof.2024.107610},
url = {https://www.sciencedirect.com/science/article/pii/S0950584924002155},
author = {Agnia Sergeyuk and Yaroslav Golubev and Timofey Bryksin and Iftekhar Ahmed}
}

@inproceedings{2022_Usability_of_code_generation_tools_Expectation_vs_Experience,
author = {Vaithilingam, Priyan and Zhang, Tianyi and Glassman, Elena L.},
title = {Expectation vs. Experience: Evaluating the Usability of Code Generation Tools Powered by Large Language Models},
year = {2022},
isbn = {9781450391566},
publisher = {Association for Computing Machinery},
address = {New York, NY, USA},
url = {https://doi-org.proxy3.library.mcgill.ca/10.1145/3491101.3519665},
doi = {10.1145/3491101.3519665},
booktitle = {Extended Abstracts of the 2022 CHI Conference on Human Factors in Computing Systems},
articleno = {332},
numpages = {7},
location = {New Orleans, LA, USA},
series = {CHI EA '22}
}

@article{Its_weird_it_knows_what_I_want,
author = {Prather, James and Reeves, Brent N. and Denny, Paul and Becker, Brett A. and Leinonen, Juho and Luxton-Reilly, Andrew and Powell, Garrett and Finnie-Ansley, James and Santos, Eddie Antonio},
title = {“It’s Weird That it Knows What I Want”: Usability and Interactions with Copilot for Novice Programmers},
year = {2023},
issue_date = {February 2024},
publisher = {Association for Computing Machinery},
address = {New York, NY, USA},
volume = {31},
number = {1},
issn = {1073-0516},
url = {https://doi-org.proxy3.library.mcgill.ca/10.1145/3617367},
doi = {10.1145/3617367},
journal = {ACM Trans. Comput.-Hum. Interact.},
month = nov,
articleno = {4},
numpages = {31}
}

@inproceedings{Zamfirescu_Pereira_2025_LLM-supported_Exploration_Program_Design_Space, series={CHI ’25},
   title={Beyond Code Generation: LLM-supported Exploration of the Program Design Space},
   url={http://dx.doi.org/10.1145/3706598.3714154},
   DOI={10.1145/3706598.3714154},
   booktitle={Proceedings of the 2025 CHI Conference on Human Factors in Computing Systems},
   publisher={ACM},
   author={Zamfirescu-Pereira, J.D. and Jun, Eunice and Terry, Michael and Yang, Qian and Hartmann, Bjoern},
   year={2025},
   month=apr, pages={1–17},
   collection={CHI ’25} }

@misc{he2025doesaiassistedcodingdeliver,
      title={Speed at the Cost of Quality: How Cursor AI Increases Short-Term Velocity and Long-Term Complexity in Open-Source Projects}, 
      author={Hao He and Courtney Miller and Shyam Agarwal and Christian Kästner and Bogdan Vasilescu},
      year={2026},
      eprint={2511.04427},
      archivePrefix={arXiv},
      primaryClass={cs.SE},
      url={https://arxiv.org/abs/2511.04427}, 
}

@inproceedings{mikami_microversioning,
author = {Mikami, Hiroaki and Sakamoto, Daisuke and Igarashi, Takeo},
title = {Micro-Versioning Tool to Support Experimentation in Exploratory Programming},
year = {2017},
isbn = {9781450346559},
publisher = {Association for Computing Machinery},
address = {New York, NY, USA},
url = {https://doi-org.proxy3.library.mcgill.ca/10.1145/3025453.3025597},
doi = {10.1145/3025453.3025597},
booktitle = {Proceedings of the 2017 CHI Conference on Human Factors in Computing Systems},
pages = {6208–6219},
numpages = {12},
location = {Denver, Colorado, USA},
series = {CHI '17}
}

@inproceedings{kery_variolite,
author = {Kery, Mary Beth and Horvath, Amber and Myers, Brad},
title = {Variolite: Supporting Exploratory Programming by Data Scientists},
year = {2017},
isbn = {9781450346559},
publisher = {Association for Computing Machinery},
address = {New York, NY, USA},
url = {https://doi-org.proxy3.library.mcgill.ca/10.1145/3025453.3025626},
doi = {10.1145/3025453.3025626},
booktitle = {Proceedings of the 2017 CHI Conference on Human Factors in Computing Systems},
pages = {1265–1276},
numpages = {12},
location = {Denver, Colorado, USA},
series = {CHI '17}
}

@inproceedings{kery_story_in_the_notebook,
author = {Kery, Mary Beth and Radensky, Marissa and Arya, Mahima and John, Bonnie E. and Myers, Brad A.},
title = {The Story in the Notebook: Exploratory Data Science using a Literate Programming Tool},
year = {2018},
isbn = {9781450356206},
publisher = {Association for Computing Machinery},
address = {New York, NY, USA},
url = {https://doi-org.proxy3.library.mcgill.ca/10.1145/3173574.3173748},
doi = {10.1145/3173574.3173748},
booktitle = {Proceedings of the 2018 CHI Conference on Human Factors in Computing Systems},
pages = {1–11},
numpages = {11},
location = {Montreal QC, Canada},
series = {CHI '18}
}

@inproceedings{IBM_Examining_Use_Impact_AI_on_Developer_Productivity_Experience,
author = {Weisz, Justin D. and Kumar, Shraddha Vijay and Muller, Michael and Browne, Karen-Ellen and Goldberg, Arielle and Heintze, Katrin Ellice and Bajpai, Shagun},
title = {Examining the Use and Impact of an AI Code Assistant on Developer Productivity and Experience in the Enterprise},
year = {2025},
isbn = {9798400713958},
publisher = {Association for Computing Machinery},
address = {New York, NY, USA},
url = {https://doi-org.proxy3.library.mcgill.ca/10.1145/3706599.3706670},
doi = {10.1145/3706599.3706670},
booktitle = {Proceedings of the Extended Abstracts of the CHI Conference on Human Factors in Computing Systems},
articleno = {673},
numpages = {13},
location = {
},
series = {CHI EA '25}
}

@book{braun_thematic_2021,
	title = {Thematic Analysis: A Practical Guide},
	isbn = {978-1-5264-1729-9},
    year = {2011},
	shorttitle = {Thematic Analysis},
	pagetotal = {369},
	publisher = {{SAGE}},
	author = {Braun, Virginia and Clarke, Victoria},
	date = {2021-10-13},
	langid = {english}
}

@article{lmerTest,
 title={lmerTest Package: Tests in Linear Mixed Effects Models},
 volume={82},
 url={https://www.jstatsoft.org/index.php/jss/article/view/v082i13},
 doi={10.18637/jss.v082.i13},
 number={13},
 journal={Journal of Statistical Software},
 author={Kuznetsova, Alexandra and Brockhoff, Per B. and Christensen, Rune H. B.},
 year={2017},
 pages={1–26}
}

@inproceedings{TheProgrammersAssistant,
author = {Ross, Steven I. and Martinez, Fernando and Houde, Stephanie and Muller, Michael and Weisz, Justin D.},
title = {The Programmer’s Assistant: Conversational Interaction with a Large Language Model for Software Development},
year = {2023},
isbn = {9798400701061},
publisher = {Association for Computing Machinery},
address = {New York, NY, USA},
url = {https://doi-org.proxy3.library.mcgill.ca/10.1145/3581641.3584037},
doi = {10.1145/3581641.3584037},
booktitle = {Proceedings of the 28th International Conference on Intelligent User Interfaces},
pages = {491–514},
numpages = {24},
location = {Sydney, NSW, Australia},
series = {IUI '23}
}

@article{BeyondCodeGenerationGPT,
author = {Khojah, Ranim and Mohamad, Mazen and Leitner, Philipp and de Oliveira Neto, Francisco Gomes},
title = {Beyond Code Generation: An Observational Study of ChatGPT Usage in Software Engineering Practice},
year = {2024},
issue_date = {July 2024},
publisher = {Association for Computing Machinery},
address = {New York, NY, USA},
volume = {1},
number = {FSE},
url = {https://doi-org.proxy3.library.mcgill.ca/10.1145/3660788},
doi = {10.1145/3660788},
journal = {Proc. ACM Softw. Eng.},
month = jul,
articleno = {81},
numpages = {22}
}

@misc{swebench,
      title={SWE-bench: Can Language Models Resolve Real-World GitHub Issues?}, 
      author={Carlos E. Jimenez and John Yang and Alexander Wettig and Shunyu Yao and Kexin Pei and Ofir Press and Karthik Narasimhan},
      year={2024},
      eprint={2310.06770},
      archivePrefix={arXiv},
      primaryClass={cs.CL},
      url={https://arxiv.org/abs/2310.06770}, 
}

@inproceedings{ivie,
author = {Yan, Litao and Hwang, Alyssa and Wu, Zhiyuan and Head, Andrew},
title = {Ivie: Lightweight Anchored Explanations of Just-Generated Code},
year = {2024},
isbn = {9798400703300},
publisher = {Association for Computing Machinery},
address = {New York, NY, USA},
url = {https://doi-org.proxy3.library.mcgill.ca/10.1145/3613904.3642239},
doi = {10.1145/3613904.3642239},
booktitle = {Proceedings of the 2024 CHI Conference on Human Factors in Computing Systems},
articleno = {140},
numpages = {15},
location = {Honolulu, HI, USA},
series = {CHI '24}
}

@inproceedings{liveProgramming,
author = {Ferdowsi, Kasra and Huang, Ruanqianqian (Lisa) and James, Michael B. and Polikarpova, Nadia and Lerner, Sorin},
title = {Validating AI-Generated Code with Live Programming},
year = {2024},
isbn = {9798400703300},
publisher = {Association for Computing Machinery},
address = {New York, NY, USA},
url = {https://doi-org.proxy3.library.mcgill.ca/10.1145/3613904.3642495},
doi = {10.1145/3613904.3642495},
booktitle = {Proceedings of the 2024 CHI Conference on Human Factors in Computing Systems},
articleno = {143},
numpages = {8},
location = {Honolulu, HI, USA},
series = {CHI '24}
}

@inproceedings{BISCUIT,
title = {BISCUIT: Scaffolding LLM-Generated Code with Ephemeral UIs in Computational Notebooks},
booktitle = {IEEE Symposium on Visual Languages and Human-Centric Computing (VL/HCC)},
author = {Ruijia Cheng and Titus Barik and Alan Leung and Fred Hohman and Jeffrey Nichols},
year = {2024},
URL = {https://arxiv.org/abs/2404.07387}
}

@inproceedings{NeedHelp,
author = {Chen, Valerie and Zhu, Alan and Zhao, Sebastian and Mozannar, Hussein and Sontag, David and Talwalkar, Ameet},
title = {Need Help? Designing Proactive AI Assistants for Programming},
year = {2025},
isbn = {9798400713941},
publisher = {Association for Computing Machinery},
address = {New York, NY, USA},
url = {https://doi-org.proxy3.library.mcgill.ca/10.1145/3706598.3714002},
doi = {10.1145/3706598.3714002},
booktitle = {Proceedings of the 2025 CHI Conference on Human Factors in Computing Systems},
articleno = {881},
numpages = {18},
location = {
},
series = {CHI '25}
}

@inproceedings{ExplainabilityScenario,
author = {Sun, Jiao and Liao, Q. Vera and Muller, Michael and Agarwal, Mayank and Houde, Stephanie and Talamadupula, Kartik and Weisz, Justin D.},
title = {Investigating Explainability of Generative AI for Code through Scenario-based Design},
year = {2022},
isbn = {9781450391443},
publisher = {Association for Computing Machinery},
address = {New York, NY, USA},
url = {https://doi-org.proxy3.library.mcgill.ca/10.1145/3490099.3511119},
doi = {10.1145/3490099.3511119},
booktitle = {Proceedings of the 27th International Conference on Intelligent User Interfaces},
pages = {212–228},
numpages = {17},
location = {Helsinki, Finland},
series = {IUI '22}
}

@inproceedings{WideningGap,
author = {Prather, James and Reeves, Brent N and Leinonen, Juho and MacNeil, Stephen and Randrianasolo, Arisoa S and Becker, Brett A. and Kimmel, Bailey and Wright, Jared and Briggs, Ben},
title = {The Widening Gap: The Benefits and Harms of Generative AI for Novice Programmers},
year = {2024},
isbn = {9798400704758},
publisher = {Association for Computing Machinery},
address = {New York, NY, USA},
url = {https://doi-org.proxy3.library.mcgill.ca/10.1145/3632620.3671116},
doi = {10.1145/3632620.3671116},
booktitle = {Proceedings of the 2024 ACM Conference on International Computing Education Research - Volume 1},
pages = {469–486},
numpages = {18},
location = {Melbourne, VIC, Australia},
series = {ICER '24}
}

@article{UncertaintyHighlighting,
author = {Vasconcelos, Helena and Bansal, Gagan and Fourney, Adam and Liao, Q. Vera and Wortman Vaughan, Jennifer},
title = {Generation Probabilities Are Not Enough: Uncertainty Highlighting in AI Code Completions},
year = {2025},
issue_date = {February 2025},
publisher = {Association for Computing Machinery},
address = {New York, NY, USA},
volume = {32},
number = {1},
issn = {1073-0516},
url = {https://doi-org.proxy3.library.mcgill.ca/10.1145/3702320},
doi = {10.1145/3702320},
journal = {ACM Trans. Comput.-Hum. Interact.},
month = apr,
articleno = {4},
numpages = {30}
}

@misc{eyeTracking,
      title={A Study on Developer Behaviors for Validating and Repairing LLM-Generated Code Using Eye Tracking and IDE Actions}, 
      author={Ningzhi Tang and Meng Chen and Zheng Ning and Aakash Bansal and Yu Huang and Collin McMillan and Toby Jia-Jun Li},
      year={2024},
      eprint={2405.16081},
      archivePrefix={arXiv},
      primaryClass={cs.SE},
      url={https://arxiv.org/abs/2405.16081}, 
}

@inproceedings{metaManager,
author = {Horvath, Amber and Macvean, Andrew and Myers, Brad A},
title = {Meta-Manager: A Tool for Collecting and Exploring Meta Information about Code},
year = {2024},
isbn = {9798400703300},
publisher = {Association for Computing Machinery},
address = {New York, NY, USA},
url = {https://doi-org.proxy3.library.mcgill.ca/10.1145/3613904.3642676},
doi = {10.1145/3613904.3642676},
booktitle = {Proceedings of the 2024 CHI Conference on Human Factors in Computing Systems},
articleno = {929},
numpages = {17},
location = {Honolulu, HI, USA},
series = {CHI '24}
}

@inproceedings{programTraces,
author = {Yan, Litao and Tao, Jeffrey and Chilton, Lydia B and Head, Andrew},
title = {Answering Developer Questions with Annotated Agent-Discovered Program Traces},
year = {2025},
isbn = {9798400720376},
publisher = {Association for Computing Machinery},
address = {New York, NY, USA},
url = {https://doi-org.proxy3.library.mcgill.ca/10.1145/3746059.3747652},
doi = {10.1145/3746059.3747652},
booktitle = {Proceedings of the 38th Annual ACM Symposium on User Interface Software and Technology},
articleno = {29},
numpages = {14},
location = {
},
series = {UIST '25}
}

@inproceedings{managingMesses,
author = {Head, Andrew and Hohman, Fred and Barik, Titus and Drucker, Steven M. and DeLine, Robert},
title = {Managing Messes in Computational Notebooks},
year = {2019},
isbn = {9781450359702},
publisher = {Association for Computing Machinery},
address = {New York, NY, USA},
url = {https://doi-org.proxy3.library.mcgill.ca/10.1145/3290605.3300500},
doi = {10.1145/3290605.3300500},
booktitle = {Proceedings of the 2019 CHI Conference on Human Factors in Computing Systems},
pages = {1–12},
numpages = {12},
location = {Glasgow, Scotland Uk},
series = {CHI '19}
}

@inproceedings{interlink,
author = {Lin, Yanna and Yang, Leni and Li, Haotian and Qu, Huamin and Moritz, Dominik},
title = {InterLink: Linking Text with Code and Output in Computational Notebooks},
year = {2025},
isbn = {9798400713941},
publisher = {Association for Computing Machinery},
address = {New York, NY, USA},
url = {https://doi-org.proxy3.library.mcgill.ca/10.1145/3706598.3714104},
doi = {10.1145/3706598.3714104},
booktitle = {Proceedings of the 2025 CHI Conference on Human Factors in Computing Systems},
articleno = {51},
numpages = {15},
location = {
},
series = {CHI '25}
}

@inproceedings{bicycle,
author = {Mendes, Wendy and Souza, Samara and De Souza, Cleidson},
title = {"You're on a bicycle with a little motor": Benefits and Challenges of Using AI Code Assistants},
year = {2024},
isbn = {9798400705335},
publisher = {Association for Computing Machinery},
address = {New York, NY, USA},
url = {https://doi-org.proxy3.library.mcgill.ca/10.1145/3641822.3641882},
doi = {10.1145/3641822.3641882},
booktitle = {Proceedings of the 2024 IEEE/ACM 17th International Conference on Cooperative and Human Aspects of Software Engineering},
pages = {144–152},
numpages = {9},
location = {Lisbon, Portugal},
series = {CHASE '24}
}

@inproceedings{designingTrust,
author = {Wang, Ruotong and Cheng, Ruijia and Ford, Denae and Zimmermann, Thomas},
title = {Investigating and Designing for Trust in AI-powered Code Generation Tools},
year = {2024},
isbn = {9798400704505},
publisher = {Association for Computing Machinery},
address = {New York, NY, USA},
url = {https://doi-org.proxy3.library.mcgill.ca/10.1145/3630106.3658984},
doi = {10.1145/3630106.3658984},
booktitle = {Proceedings of the 2024 ACM Conference on Fairness, Accountability, and Transparency},
pages = {1475–1493},
numpages = {19},
location = {Rio de Janeiro, Brazil},
series = {FAccT '24}
}

@inproceedings{coladder,
author = {Yen, Ryan and Zhu, Jiawen Stefanie and Suh, Sangho and Xia, Haijun and Zhao, Jian},
title = {CoLadder: Manipulating Code Generation via Multi-Level Blocks},
year = {2024},
isbn = {9798400706288},
publisher = {Association for Computing Machinery},
address = {New York, NY, USA},
url = {https://doi-org.proxy3.library.mcgill.ca/10.1145/3654777.3676357},
doi = {10.1145/3654777.3676357},
booktitle = {Proceedings of the 37th Annual ACM Symposium on User Interface Software and Technology},
articleno = {11},
numpages = {20},
location = {Pittsburgh, PA, USA},
series = {UIST '24}
}

@inproceedings{10.1145/2047196.2047225,
author = {Karrer, Thorsten and Kr\"{a}mer, Jan-Peter and Diehl, Jonathan and Hartmann, Bj\"{o}rn and Borchers, Jan},
title = {Stacksplorer: call graph navigation helps increasing code maintenance efficiency},
year = {2011},
isbn = {9781450307161},
publisher = {Association for Computing Machinery},
address = {New York, NY, USA},
doi = {10.1145/2047196.2047225},
booktitle = {Proceedings of the 24th Annual ACM Symposium on User Interface Software and Technology},
pages = {217–224},
numpages = {8},
location = {Santa Barbara, California, USA},
series = {UIST '11}
}

@inproceedings{10.1145/3025453.3025645,
    author = {Henley, Austin Z. and Fleming, Scott D. and Luong, Maria V.},
    title = {Toward Principles for the Design of Navigation Affordances in Code Editors: An Empirical Investigation},
    year = {2017},
    isbn = {9781450346559},
    publisher = {Association for Computing Machinery},
    address = {New York, NY, USA},
    doi = {10.1145/3025453.3025645},
    booktitle = {Proceedings of the 2017 CHI Conference on Human Factors in Computing Systems},
    pages = {5690–5702},
    numpages = {13},
    location = {Denver, Colorado, USA},
    series = {CHI '17}
    }

@inproceedings{10.1145/3540250.3549084,
author = {Peitek, Norman and Bergum, Annabelle and Rekrut, Maurice and Mucke, Jonas and Nadig, Matthias and Parnin, Chris and Siegmund, Janet and Apel, Sven},
title = {Correlates of programmer efficacy and their link to experience: a combined EEG and eye-tracking study},
year = {2022},
isbn = {9781450394130},
publisher = {Association for Computing Machinery},
address = {New York, NY, USA},
doi = {10.1145/3540250.3549084},
booktitle = {Proceedings of the 30th ACM Joint European Software Engineering Conference and Symposium on the Foundations of Software Engineering},
pages = {120–131},
numpages = {12},
location = {Singapore, Singapore},
series = {ESEC/FSE 2022}
}

@inproceedings{10.1145/1287624.1287673,
author = {Fritz, Thomas and Murphy, Gail C. and Hill, Emily},
title = {Does a programmer's activity indicate knowledge of code?},
year = {2007},
isbn = {9781595938114},
publisher = {Association for Computing Machinery},
address = {New York, NY, USA},
doi = {10.1145/1287624.1287673},
booktitle = {Proceedings of the the 6th Joint Meeting of the European Software Engineering Conference and the ACM SIGSOFT Symposium on The Foundations of Software Engineering},
pages = {341–350},
numpages = {10},
location = {Dubrovnik, Croatia},
series = {ESEC-FSE '07}
}

@article{TICODER,
author = {Fakhoury, Sarah and Naik, Aaditya and Sakkas, Georgios and Chakraborty, Saikat and Lahiri, Shuvendu K.},
title = {LLM-Based Test-Driven Interactive Code Generation: User Study and Empirical Evaluation},
year = {2024},
issue_date = {Sept. 2024},
publisher = {IEEE Press},
volume = {50},
number = {9},
issn = {0098-5589},
url = {https://doi.org/10.1109/TSE.2024.3428972},
doi = {10.1109/TSE.2024.3428972},
journal = {IEEE Trans. Softw. Eng.},
month = sep,
pages = {2254–2268},
numpages = {15}
}

@inproceedings{Arawjo2024ChainForge,
author = {Arawjo, Ian and Swoopes, Chelse and Vaithilingam, Priyan and Wattenberg, Martin and Glassman, Elena L.},
title = {ChainForge: A Visual Toolkit for Prompt Engineering and LLM Hypothesis Testing},
year = {2024},
isbn = {9798400703300},
publisher = {Association for Computing Machinery},
address = {New York, NY, USA},
url = {https://doi.org/10.1145/3613904.3642016},
doi = {10.1145/3613904.3642016},
booktitle = {Proceedings of the 2024 CHI Conference on Human Factors in Computing Systems},
articleno = {304},
numpages = {18},
location = {Honolulu, HI, USA},
series = {CHI '24}
}

@inproceedings{Coscia2025OnGoal,
author = {Coscia, Adam J and Guo, Shunan and Koh, Eunyee and Endert, Alex},
title = {OnGoal: Tracking and Visualizing Conversational Goals in Multi-Turn Dialogue with Large Language Models},
year = {2025},
isbn = {9798400720376},
publisher = {Association for Computing Machinery},
address = {New York, NY, USA},
url = {https://doi.org/10.1145/3746059.3747746},
doi = {10.1145/3746059.3747746},
booktitle = {Proceedings of the 38th Annual ACM Symposium on User Interface Software and Technology},
articleno = {208},
numpages = {18},
location = {
},
series = {UIST '25}
}

@article{demandChars,
author = {Iarygina, Olga and Hornb\ae{}k, Kasper and Mottelson, Aske},
title = {Demand characteristics in human–computer experiments},
year = {2025},
issue_date = {Jan 2025},
publisher = {Academic Press, Inc.},
address = {USA},
volume = {193},
number = {C},
issn = {1071-5819},
url = {https://doi-org.proxy3.library.mcgill.ca/10.1016/j.ijhcs.2024.103379},
doi = {10.1016/j.ijhcs.2024.103379},
journal = {Int. J. Hum.-Comput. Stud.},
month = jan,
numpages = {14}
}

@inproceedings{PerfectionNotRequired,
author = {Weisz, Justin D. and Muller, Michael and Houde, Stephanie and Richards, John and Ross, Steven I. and Martinez, Fernando and Agarwal, Mayank and Talamadupula, Kartik},
title = {Perfection Not Required? Human-AI Partnerships in Code Translation},
year = {2021},
isbn = {9781450380171},
publisher = {Association for Computing Machinery},
address = {New York, NY, USA},
url = {https://doi-org.proxy3.library.mcgill.ca/10.1145/3397481.3450656},
doi = {10.1145/3397481.3450656},
booktitle = {Proceedings of the 26th International Conference on Intelligent User Interfaces},
pages = {402–412},
numpages = {11},
location = {College Station, TX, USA},
series = {IUI '21}
}

@online{StackOverflowSurvey2025,
  author       = {Stack Overflow},
  title        = {Developers Remain Willing, but Reluctant, to Use AI: The 2025 Developer Survey Results Are Here},
  year         = {2025},
  month        = dec,
  day          = {29},
  url          = {https://stackoverflow.blog/2025/12/29/developers-remain-willing-but-reluctant-to-use-ai-the-2025-developer-survey-results-are-here/},
  urldate      = {2026-03-29},
  organization = {Stack Overflow Blog}
}

@misc{mao2024multiuserchatassistantmuca,
      title={Multi-User Chat Assistant (MUCA): a Framework Using LLMs to Facilitate Group Conversations}, 
      author={Manqing Mao and Paishun Ting and Yijian Xiang and Mingyang Xu and Julia Chen and Jianzhe Lin},
      year={2024},
      eprint={2401.04883},
      archivePrefix={arXiv},
      primaryClass={cs.CL},
      url={https://arxiv.org/abs/2401.04883}, 
}

@misc{tbs2025directive,
  author = "{Canada. Treasury Board}",
  title = {Directive on {A}utomated {D}ecision-{M}aking},
  howpublished = {Government of Canada},
  year = {2021},
  url = {https://www.tbs-sct.canada.ca/pol/doc-eng.aspx?id=32592},
  note = {Accessed: 2024-03-29}
}

@misc{eu2024aiact,
  author = {{European Parliament and the Council of the European Union}},
  title = {Regulation ({EU}) 2024/1689 of the {E}uropean {P}arliament and of the {C}ouncil of 13 {J}une 2024 laying down harmonised rules on artificial intelligence ({A}rtificial {I}ntelligence {A}ct)},
  howpublished = {Official Journal of the European Union, L 2024/1689},
  year = {2024},
  url = {http://data.europa.eu/eli/reg/2024/1689/oj},
  note = {Accessed: 2026-03-29}
}

@book{montgomeryDesign,
    author = {Montgomery, Douglas C.},
    title = {Design and Analysis of Experiments},
    year = {2006},
    isbn = {0470088109},
    publisher = {John Wiley \& Sons, Inc.},
    address = {Hoboken, NJ, USA}
}

@article{agenticCoding,
author = {Watanabe, Miku and Li, Hao and Kashiwa, Yutaro and Reid, Brittany and Iida, Hajimu and Hassan, Ahmed E.},
title = {On the Use of Agentic Coding: An Empirical Study of Pull Requests on GitHub},
year = {2026},
publisher = {Association for Computing Machinery},
address = {New York, NY, USA},
issn = {1049-331X},
url = {https://doi-org.proxy3.library.mcgill.ca/10.1145/3798166},
doi = {10.1145/3798166},
note = {Just Accepted},
journal = {ACM Trans. Softw. Eng. Methodol.},
month = mar
}

@misc{clark1991grounding,
  address = {Washington,  DC,  US},
  author = {Clark, Herbert H. and Brennan, Susan E.},
  booktitle = {Perspectives on socially shared cognition.},
  doi = {10.1037/10096-006},
  isbn = {1557981213},
  pages = {127-149},
  publisher = {American Psychological Association},
  refid = {1991-98452-006},
  title = {Grounding in communication},
  url = {https://psycnet.apa.org/record/1991-98452-006},
  year = 1991
}

@article{Olson01092000,
author = {Gary M. Olson and Judith S. Olson},
title = {Distance Matters},
journal = {Human–Computer Interaction},
volume = {15},
number = {2-3},
pages = {139--178},
year = {2000},
publisher = {Taylor \& Francis},
doi = {10.1207/S15327051HCI1523\_4},
URL = { https://doi.org/10.1207/S15327051HCI1523_4},
}

@article{AccountabilityInCodeReview,
author = {Alami, Adam and Jensen, Victor and Ernst, Neil},
title = {Accountability in Code Review: The Role of Intrinsic Drivers and the Impact of LLMs},
year = {2025},
issue_date = {November 2025},
publisher = {Association for Computing Machinery},
address = {New York, NY, USA},
volume = {34},
number = {8},
issn = {1049-331X},
url = {https://doi-org.proxy3.library.mcgill.ca/10.1145/3721127},
doi = {10.1145/3721127},
journal = {ACM Trans. Softw. Eng. Methodol.},
month = oct,
articleno = {233},
numpages = {44}
}

@inproceedings{DBLP:conf/kbse/Jiang19,
  author       = {Shuyao Jiang},
  title        = {Boosting Neural Commit Message Generation with Code Semantic Analysis},
  booktitle    = {34th {IEEE/ACM} International Conference on Automated Software Engineering,
                  {ASE} 2019, San Diego, CA, USA, November 11-15, 2019},
  pages        = {1280--1282},
  publisher    = {{IEEE}},
  year         = {2019},
  url          = {https://doi.org/10.1109/ASE.2019.00162},
  doi          = {10.1109/ASE.2019.00162},
}

@inproceedings{DBLP:conf/icse/ZhouVK20a,
  author       = {Shurui Zhou and
                  Bogdan Vasilescu and
                  Christian K{\"{a}}stner},
  title        = {How Has Forking Changed in the Last 20 Years?: A Study of Hard Forks on Github},
  booktitle    = {{ICSE} '20: 42nd International Conference on Software Engineering,
                  Companion Volume, Seoul, South Korea, 27 June - 19 July, 2020},
  pages        = {268--269},
  publisher    = {{ACM}},
  year         = {2020},
  url          = {https://doi.org/10.1145/3377812.3390911},
  doi          = {10.1145/3377812.3390911},
}

@incollection{guo2025natural,
  title={Natural language processing for requirements traceability},
  author={Guo, Jin LC and Stegh{\"o}fer, Jan-Philipp and Vogelsang, Andreas and Cleland-Huang, Jane},
  booktitle={Handbook on Natural Language Processing for Requirements Engineering},
  pages={89--116},
  year={2025},
  publisher={Springer}
}

@inproceedings{DBLP:conf/icse/0004CC17,
  author       = {Jin Guo and
                  Jinghui Cheng and
                  Jane Cleland{-}Huang},
  title        = {Semantically enhanced software traceability using deep learning techniques},
  booktitle    = {Proceedings of the 39th International Conference on Software Engineering,
                  {ICSE} 2017, Buenos Aires, Argentina, May 20-28, 2017},
  pages        = {3--14},
  publisher    = {{IEEE} / {ACM}},
  year         = {2017},
  url          = {https://doi.org/10.1109/ICSE.2017.9},
  doi          = {10.1109/ICSE.2017.9},
}
\bibliographystyle{ACM-Reference-Format}

\appendix
\section{User Study Protocol}
\label{sec:appendix}
\begin{enumerate}
    \item Provide the participant with the appropriate tutorial for their first task, based on the version of the tool they will use.
    \item Ask the participant to complete the first task.
    \item Have the participant complete the post-task questionnaire for the first task.
    \item Provide the participant with the appropriate tutorial for their second task, based on the version of the tool they will use.
    \item Ask the participant to complete the second task.
    \item Have the participant complete the post-task questionnaire for the second task.
    \item Ask the participant the post-study questions.
\end{enumerate}

\subsection{Tutorials}
\subsubsection{Control}
\begin{enumerate}
\item Have the participant open the Tutorial repository, using the Control version of the tool.
\item Ask the participant to prompt the assistant to add a section about Python to the README file.
\item Have the participant apply the code edit.
\item Ask the participant to use the edit message feature to revert the changes and instead prompt the assistant to add a section about Java to the README file.
\item Have the participant apply the code edit.
\item Have the participant open a new chat and send a message.
\item Have the participant switch back to their original chat tab.
\end{enumerate}

\subsubsection{\toolname{}}
\begin{enumerate}
\item Have the participant open the Tutorial repository, using \toolname{}.
\item Ask the participant to prompt the assistant to add a section about Python to the README file.
\item Have the participant apply the code edit.
\item Ask the participant to use the graph to revert sequentially to the origin node.
\item Ask the participant to prompt the assistant to add a section about Java to the README file.
\item Have the participant apply the code edit.
\item Have the participant make a small manual change and create a checkpoint.
\item Have the participant compare the nodes at the two ends of the Python and Java branches.
\item Have the participant merge the branches together.
\item Have the participant use Review Mode to see the origin of each code block.
\item Have the participant open a new chat.
\item Have the participant ask the assistant to explain the evolution of the codebase and include the development graph in the context.
\end{enumerate}

\subsection{Tasks}

\subsubsection{Weather Dashboard Task}
You have been given a basic weather dashboard starter code. Your task is to transform it into a more polished and better-equipped application within 30 minutes. By the end of the task, you will need to write a short report summarizing what features you have added and how they work. When writing the report, you can review your code and prior AI interactions, but you cannot use the assistant to generate the report for you. You will have 25 minutes for the coding portion and 5 minutes at the end to write up the report.

In case you need some inspiration, here are some potential features you can choose to implement, but you have the freedom to decide what features to implement and how to implement them: 
\begin{itemize}
\item Improve the design with additional styling and layout adjustments. 
\item Add a unit toggle for Celsius/Fahrenheit and km/h vs mph. 
\item Let users save favourite cities whose weather is always shown. 
\item Implement autocomplete for the city search bar. 
\item Display dynamic backgrounds that reflect the current weather or time of day. 
\item Add a graph or chart to show hourly weather changes.
\end{itemize}
For the task, you are allowed to use the tool, your browser to check the dashboard, and git for versioning.

\subsubsection{Task Dashboard Task}
You have been given a basic task management dashboard starter code. Your task is to transform it into a more polished and better-equipped application within 30 minutes. By the end of the task, you will need to write a short report summarizing what features you have added and how they work. When writing the report, you can review your code and prior AI interactions, but you cannot use the assistant to generate the report for you. You will have 25 minutes for the coding portion and 5 minutes at the end to write up the report.

In case you need some inspiration, here are some potential features you can choose to implement, but you have the freedom to decide what features to implement and how to implement them:
\begin{itemize}
\item Improve the design with additional styling and layout adjustments.
\item Add options to create, edit, and delete tasks.
\item Let users organize tasks by categories, tags, or priority levels.
\item Implement due dates with visual indicators for upcoming or overdue tasks.
\item Provide a search or filter option to quickly find specific tasks.
\item Add a dashboard summary showing task counts (e.g., completed, pending, overdue).
\end{itemize}
For the task, you are allowed to use the tool, your browser to check the dashboard, and git for versioning.

\subsection{Post-Task Questionnaire}
\textbf{Workload (NASA-TLX)}

\begin{itemize}
    \item How mentally demanding was the task? (1–7): Very Low – Very High
    \item How successful were you in accomplishing what you were asked to do? (1–7): Perfect – Failure
    \item How hard did you have to work to accomplish your level of performance? (1–7): Very Low – Very High
    \item How insecure, discouraged, irritated, stressed, and annoyed were you? (1–7): Very Low – Very High
\end{itemize}

\textbf{Context Management}

\begin{itemize}
    \item How easy or difficult was it to review the progression of the chat history with the LLM? (1–7): Very Easy – Very Difficult
    \item How easy or difficult was it to modify or refine prompts based on initial responses? (1–7): Very Easy – Very Difficult
    \item How easy or difficult was it to give the AI assistant the context of your development progress? (1–7): Very Easy – Very Difficult
\end{itemize}

\textbf{Exploration}

\begin{itemize}
    \item How easy or difficult was it to revert incorrect changes? (1–7): Very Easy – Very Difficult
    \item How easy or difficult was it to try alternative solutions? (1–7): Very Easy – Very Difficult
    \item How easy or difficult was it to compare alternate solutions? (1–7): Very Easy – Very Difficult
    \item How easy or difficult was it to combine alternate solutions together? (1–7): Very Easy – Very Difficult
\end{itemize}

\textbf{Provenance}

\begin{itemize}
    \item How easy or difficult was it to trace where a particular code change came from (e.g., AI suggestion vs. your own edit)? (1–7): Very Easy – Very Difficult
    \item How aware did you feel of the changes being made in the codebase? (1–7): Very Aware – Very Unaware
    \item How easy or difficult was it to review the progression of your codebase? (1–7): Very Easy – Very Difficult
\end{itemize}

\subsection{Post-Study Interview}
Now that you have completed both tasks, let’s compare your experiences.
\begin{itemize}
\item Did you find there were any differences in your programming approach between the two tasks? If so, what were those differences, and what caused them?
\item Did you notice any benefits when using one tool compared to the other? If so, what were they, and what contributed to them?
\item Did you notice any drawbacks when using one tool compared to the other? If so, what were they, and what contributed to them?
\item When performing the task with each tool, how would you describe your overall workload? Did one feel more mentally or cognitively demanding than the other, and what influenced that perception?
\end{itemize}

\end{document}